\documentclass[aps,
               twocolumn,
               groupedaddress,
               prab,
               floatfix
               ]{revtex4-1}

\AtBeginDocument{
    \heavyrulewidth =.08em
    \lightrulewidth =.05em
    \cmidrulewidth  =.03em
    \belowrulesep   =.65ex
    \belowbottomsep =0pt
    \aboverulesep   =.4ex
    \abovetopsep    =0pt
    \cmidrulesep    =\doublerulesep
    \cmidrulekern   =.5em
    \defaultaddspace=.5em
}

\makeatletter
\def\ps@pprintTitle{%
    \let\@oddhead\@empty
    \let\@evenhead\@empty
    \def\@oddfoot{\reset@font\hfil\thepage\hfil}
    \let\@evenfoot\@oddfoot
}
\makeatother

\usepackage{amsmath, amssymb}
\usepackage{mathtools}
\usepackage{siunitx}
\usepackage{xcolor}
\usepackage{listings}
\usepackage{forloop}
\usepackage{booktabs}

\usepackage[english]{babel}

\usepackage{multirow}

\usepackage{pifont}
\usepackage{tikz}
\usetikzlibrary{arrows}
\usetikzlibrary{calc}
\usetikzlibrary{intersections, quotes,angles,positioning}

\colorlet{RED}{red}

\newcommand{\Figref}[1]{Fig.~\ref{#1}}
\newcommand{\Tabref}[1]{Tab.~\ref{#1}}
\newcommand{\Eqref}[1]{Eq.~\eqref{#1}}

\newcommand{\Secref}[1]{Sec.~\ref{#1}}

\newcommand{\opal}{\textsc{OPAL}}

\newcommand{\nsga}{\textsc{NSGA-II}}
\newcommand{\python}{\textsc{Python}}
\newcommand{\scipy}{\textsc{SciPy}}

\newcommand{\xvec}{\mathbf{x}}
\newcommand{\vvec}{\mathbf{v}}
\newcommand{\Evec}{\mathbf{E}}
\newcommand{\Bvec}{\mathbf{B}}

\newcommand*\circled[1]{\tikz[baseline=(char.base)]{
            \node[shape=circle,draw,inner sep=1.5pt, line width=1.3pt] (char) {#1};}}

\begin{document}
    \begin{abstract}
    The usage of numerical models to study the evolution of particle beams is
    an essential step in the design process of particle accelerators.\ However,
    uncertainties of input quantities such as beam energy and magnetic field
    lead to simulation results that do not fully agree with measurements, hence the
    final machine will behave slightly differently than the simulations.\ In
    case of cyclotrons such discrepancies affect the overall turn pattern or may
    even alter the number of turns in the machine.\ Inaccuracies at the PSI
    Ring cyclotron facility that may harm the isochronism are compensated by
    additional magnetic fields provided by 18 trim coils.\ These are often absent
    from simulations or their implementation is very simplistic.\ In this paper a newly
    developed realistic trim coil model within the particle accelerator framework
    \opal{} is presented that was used to match the turn pattern of the PSI
    Ring cyclotron.\ Due to the high-dimensional search space consisting of
    \num{48} design variables (simulation input parameters) and \num{182} objectives
    (i.e.\ turns) simulation and measurement cannot be matched in a straightforward manner.\
    Instead, an evolutionary multi-objective optimization with a population size of more than
    \num{8000} individuals per generation together with a local search
    approach were applied that reduced the maximum absolute error to \SI{4.54}{mm}
    over all \num{182} turns.
\end{abstract}

    \title{Matching of turn pattern measurements for cyclotrons using multi-objective
optimization}
    
    \author{Matthias Frey}
    \email[]{matthias.frey@psi.ch}
    \affiliation{Paul Scherrer Institut, CH-5232 Villigen PSI, Switzerland}
    
    \author{Jochem Snuverink}
    \email[]{jochem.snuverink@psi.ch}
    \affiliation{Paul Scherrer Institut, CH-5232 Villigen PSI, Switzerland}
    
    \author{Christian Baumgarten}
    \email[]{christian.baumgarten@psi.ch}
    \affiliation{Paul Scherrer Institut, CH-5232 Villigen PSI, Switzerland}
    
    \author{Andreas Adelmann}
    \email[]{andreas.adelmann@psi.ch}
    \affiliation{Paul Scherrer Institut, CH-5232 Villigen PSI, Switzerland}

    \maketitle
    
    \section{Introduction}
The PSI Ring cyclotron was commissioned in the mid-1970s
and has been in user operation since.\ It has a long history of upgrades and
improvements that made it possible to operate the machine with currents
of up to \SI{2.4}{mA}, a figure exceeding the design specification by a factor of \num{24}.\
But operation and development of accelerators over several decades is
challenging in many ways.\ Keeping documentation up-to-date has been proven
very challenging and in some cases even impossible.\
Beamlines or insertion devices have been modified or replaced,
sputtering processes change the form of collimators and apertures.\
However, a post commissioning survey of possibly activated accelerator
components is difficult and risky as it requires a partial
or total disassembly, hence components may not be accessible with reasonable effort.\
Thus, it is no surprise that the work to improve, optimise, replace or test
models of accelerators continues even after decades of successful operation.\

Here we report our efforts to model and fit the turn pattern of the PSI Ring 
cyclotron, for which accurate magnetic field data of the trim coil fields
are not available.\ Instead the average field profiles of the trim coils are 
derived from measurements of beam phase shifts.\
If a local field change by some inner trim coil causes a local radial shift of 
some turn, this subsequently leads, in combination with a slight difference in 
betatron tune, -phase and -amplitude, to a significant difference in the overall 
turn pattern.\ This becomes more and more significant from turn to turn.\ Hence,
the overall fit should be most sensitive to the innermost trim coils.\

Besides the contributions of the trim coils to the total magnetic field, 
the accuracy of the voltage profiles of the RF resonators plays a crucial 
role for the exact form of the turn pattern too.\ The PSI Ring cyclotron is equipped
with 5 RF cavities, i.e.\ 4 main cavities and a third harmonic flattop cavity
(see Fig.~\ref{fig:psi_ring_cyclotron}).\ While all main cavities could, 
in principle, have the same field profiles, they are not always operated
at the exact same voltage.\ Furthermore, due to the sheer size of the Ring 
cyclotron, the exact position of the cavities might slightly differ from one 
to another.\ Hence, a fit of the turn pattern in the Ring must very likely allow 
for (small) variations of the positions and voltages of the cavities. 

Various computer codes are able to compute turn patterns of cyclotrons.\
A survey of the most common cyclotron codes is given in \cite{PhysRevAccelBeams.20.124801}.\
A first step towards a realistic numerical model of the PSI Ring cyclotron
using \opal{} was taken in \cite{PhysRevSTAB.14.054402}, but only the last few turns
before extraction and one of the 18 trim coils were included.\
Another preceding study \cite{Pogue:IPAC2017-THPAB077} has shown, however,
that it is a challenging task to match all \num{182}~turns and likely requires 
more free parameters with at least all trim coil amplitudes, but also cavity voltages
and possibly cavity alignment errors.\

Due to the large design and objective spaces a simple parameter tweaking by hand is infeasible.\
A remedy is the use of global optimization techniques. In this paper
an available framework of an evolutionary algorithm is applied that is complemented
with a local search.\
Several papers such as \cite{HUANG201448, PANG2014124, YANG200950, HUSAIN2018151,
PhysRevSTAB.8.034202, PhysRevSTAB.14.054001, PhysRevSTAB.16.010101} already showed
the successful application of evolutionary algorithms like particle swarm,
differential evolution or \nsga{} \cite{996017} in connection with particle
accelerator modelling.

In this paper a general trim coil model is presented that is
integrated into the particle accelerator framework \opal{} \cite{opal:1}.\ It allows
a more realistic description of the magnetic fields based on measured data.\ Together
with the built-in multi-objective genetic algorithm (MOGA) and a local search it was possible
to match all turns of the PSI Ring cyclotron to a maximum absolute error of
\SI{4.54}{mm}.\ With the exception of \cite{Baumgarten:Cyclotrons2016-THP16, Pogue:IPAC2017-THPAB077}
the authors aren't aware of any paper that tries to match the measured turn pattern
with simulation in context of cyclotrons.\

The paper is structured as follows: In \Secref{sec:psi_ring_cyclotron} the aforementioned
cyclotron is described and the new trim coil model is explained before the next
chapter discusses its modelling and implementation within \opal.\ The results of
both approaches, i.e.\ local search and MOGA, are shown in
\Secref{sec:turn_pattern_matching} with a closer discussion in \Secref{sec:discussion}.\
Final remarks and a conclusion are gathered in the last section.

    \section{PSI Ring Cyclotron}
\label{sec:psi_ring_cyclotron}
\begin{figure}[htp]
    \centering
    \includegraphics[width=\columnwidth]{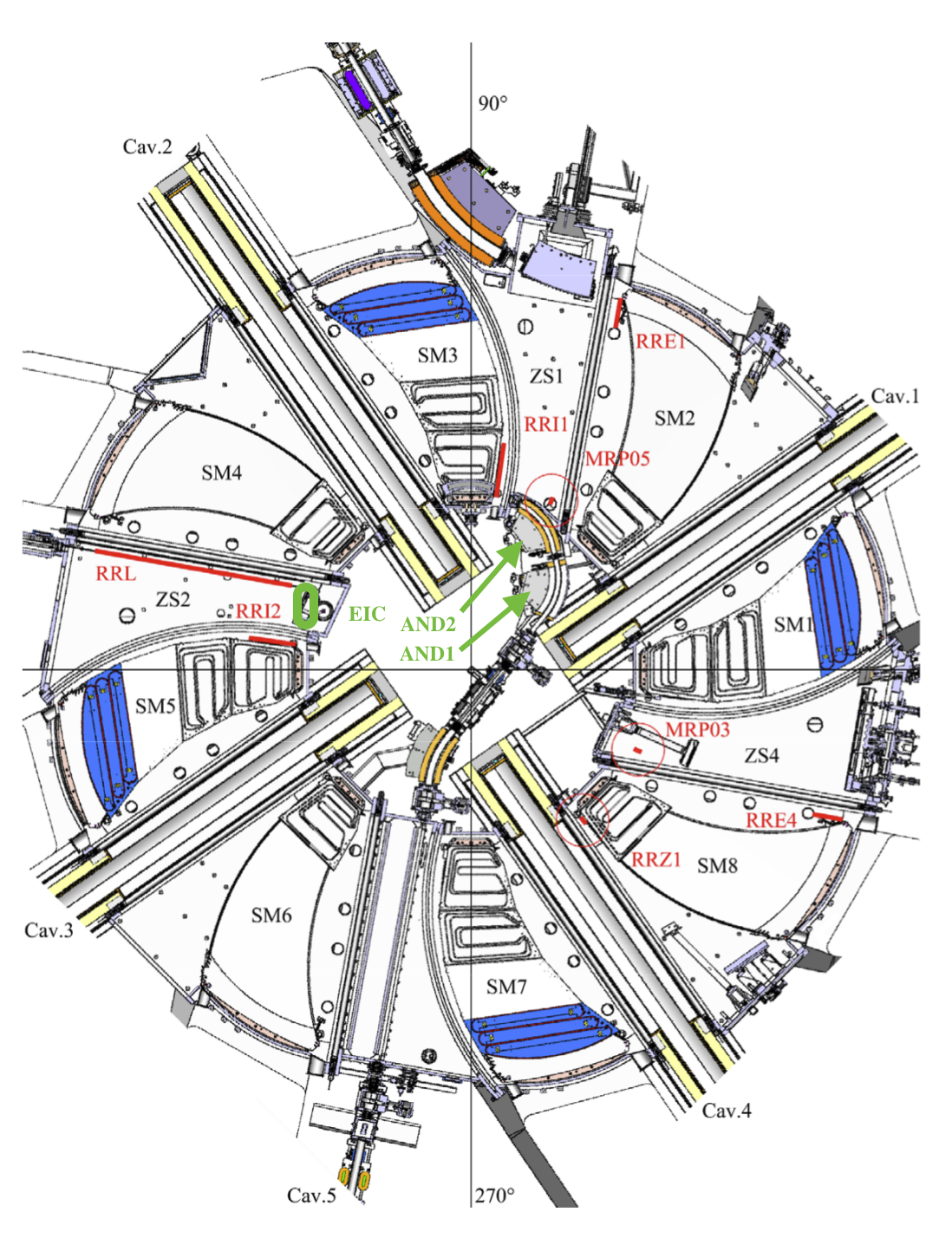}
    \caption{Plan of the PSI Ring cyclotron. Probes and monitors marked in red \cite{doelling:report}.
    Injection elements are marked in green.}
    \label{fig:psi_ring_cyclotron}
\end{figure}

\Figref{fig:psi_ring_cyclotron} shows the eight sector (SM1 - SM8)
Ring cyclotron at PSI that is the last accelerating stage of the HIPA (High Intensity Proton
Accelerator) facility accelerating routinely \SI{2.2}{mA} (max.\ \SI{2.4}{mA}) proton beams with the four main
cavities (Cav.\ \num{1} - \num{4}) and one flat top cavity (Cav.\ 5) at
\SI[round-precision=2]{50.65}{MHz} from \SIrange{72}{590}{MeV}.\ A beam is injected at an azimuth
of \SI{110}{\degree} and a radius around \SI{2}{m}.\ After typically \num{182}~turns
it is extracted and guided to several targets to produce either muons
(through pion decay) or neutrons.

\subsection{Radial probes}
The cyclotron is equipped with a total of \num{5}
wire probes that enable beam profiles~\cite{doelling:report}.\ 
However, here we use only data of the probes RRI2 (turns \numrange{1}{16}) 
and RRL (turns \numrange{9}{182}).\ 

\begin{figure}[htp]
    \centering
    \begin{tikzpicture}[axis/.style={very thick, ->, >=stealth'},
                        myarrow/.style={thick, ->, >=stealth'},
                        extended line/.style={shorten >=-#1,shorten <=-#1},
                        extended line/.default=1cm,
                        one end extended/.style={shorten >=-#1},
                        one end extended/.default=1cm,
                        dot/.style={circle,inner sep=1.5pt,fill}, scale=0.75]
    
    \node[dot,label={[below left]$(0, 0)$}] (center) at (0, 0) {};
    \node[] (a) at (3.5, -2) {};
    \node[] (p) at (4, 3.5) {};
    \node[] (c) at (5.5, 1) {};
    
    \draw[axis, name path=xaxis] (center) -- (4,0) node(xline)[right] {$x, 0\si{\degree}$};
    \draw[axis, name path=yaxis] (center) -- (0,4) node(yline)[above] {$y$};
    
    \node[dot, label={[below left]$(x_0, y_0)$}] (b) at ($(center)!(p)!(a)$) {};
    \draw[myarrow, red, name path=wire] (b) -- (p)
        node[right]{$s$}
        node[dot, pos=0.6, label={[below right]$s_{1}$}] {}
        node[dot, pos=0.8, label={[below right]$s_{2}$}] {};
    
    \draw[thick, blue, dashed] (center) -- (b) node[midway, below left] {$a$};
    
    \draw[myarrow, blue, name path=impact] (b) -- (a) {};
    
    \draw[draw=none] (a) -- (b) -- (p) pic [draw, thick, <->, >=stealth', angle radius=12mm, "$90\si{\degree}$"]
        {angle = a--b--p};
        
    \path [name intersections={of=xaxis and wire,by=inter}];
    \draw[draw=none] (xline) -- (inter) -- (p) pic [draw, thick, <->, >=stealth', angle radius=12mm, "$\varphi$"]
        {angle = xline--inter--p};
    \end{tikzpicture}
    \caption{Mathematical description of probe positioning.\ The probe orientation
    is given by $s$ with begin $s_1$ and end $s_2$.\ The offset of the probe from the
    machine center is indicated by the dashed line with symbol $a$.\ \cite{doelling:report}}
    \label{fig:probe_positioning}
\end{figure}
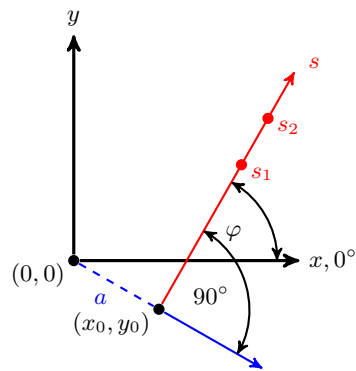

The beam profiles are obtained by measuring the current of the wire while it
moves radially through the median plane and crosses subsequently multiple turns.\  
The step width of the RRI2 and RRL probes are \SI{0.05}{mm} and \SI{0.5}{mm} respectively.\
The RRL probe has a single vertical carbon wire, while the RRI2 probe has \num{3}~carbon wires,
\num{2}~crossed and \num{1}~vertical, the combination of which enables to obtain
information about shape and vertical position of the beam.\
Here we use exclusively information of the vertical wires.\
Examples of the (normalized) profile measurement for the RRI2 and RRL probes are given in
\Figref{fig:histogram_rri2} and \Figref{fig:histogram_rrl1} respectively.\
The wire position, relative to the machine center at $(0,0)$, is described by
(cf.\ \Figref{fig:probe_positioning})
\begin{equation}
    \begin{pmatrix}
        x \\ y
    \end{pmatrix}
    = s\cdot
    \begin{pmatrix*}[r]
        \cos(\varphi) \\
        \sin(\varphi)
    \end{pmatrix*}
    +
    \begin{pmatrix}
        x_{0} \\ y_{0}
    \end{pmatrix}
    \label{eq:probe_positioning}
\end{equation}
with azimuth $\varphi$ and $s\in\left[s_1, s_2\right]$ and offset to the origin
\begin{equation*}
    \begin{pmatrix}
        x_{0} \\ y_{0}
    \end{pmatrix}
    = a\cdot
    \begin{pmatrix*}[r]
        \sin(\varphi) \\
        -\cos(\varphi)
    \end{pmatrix*}
\end{equation*}
where $a\in\mathbb{R}$.

\subsection{Peak detection of probe measurement}

\begin{figure}[htp]
    \centering
    \includegraphics[width=\columnwidth]{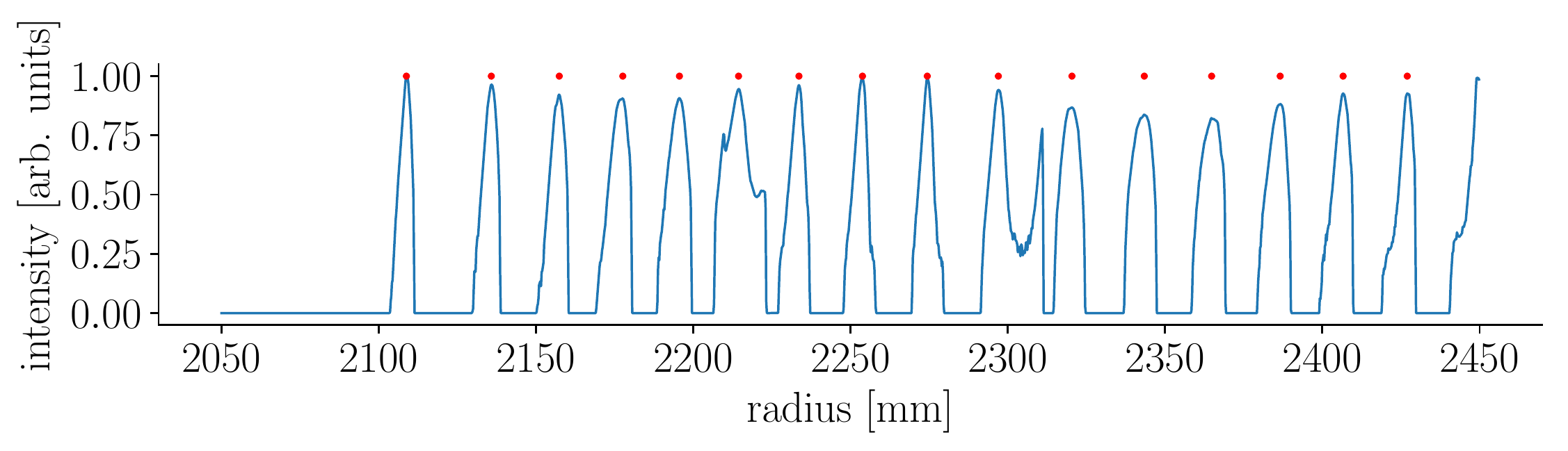}
    \caption{Histogram of the probe RRI2 measurement.\ The intensity is normalized.\
    The red dots mark detected peaks.}
    \label{fig:histogram_rri2}
\end{figure}

\begin{figure}[htp]
    \centering
    \includegraphics[width=\columnwidth]{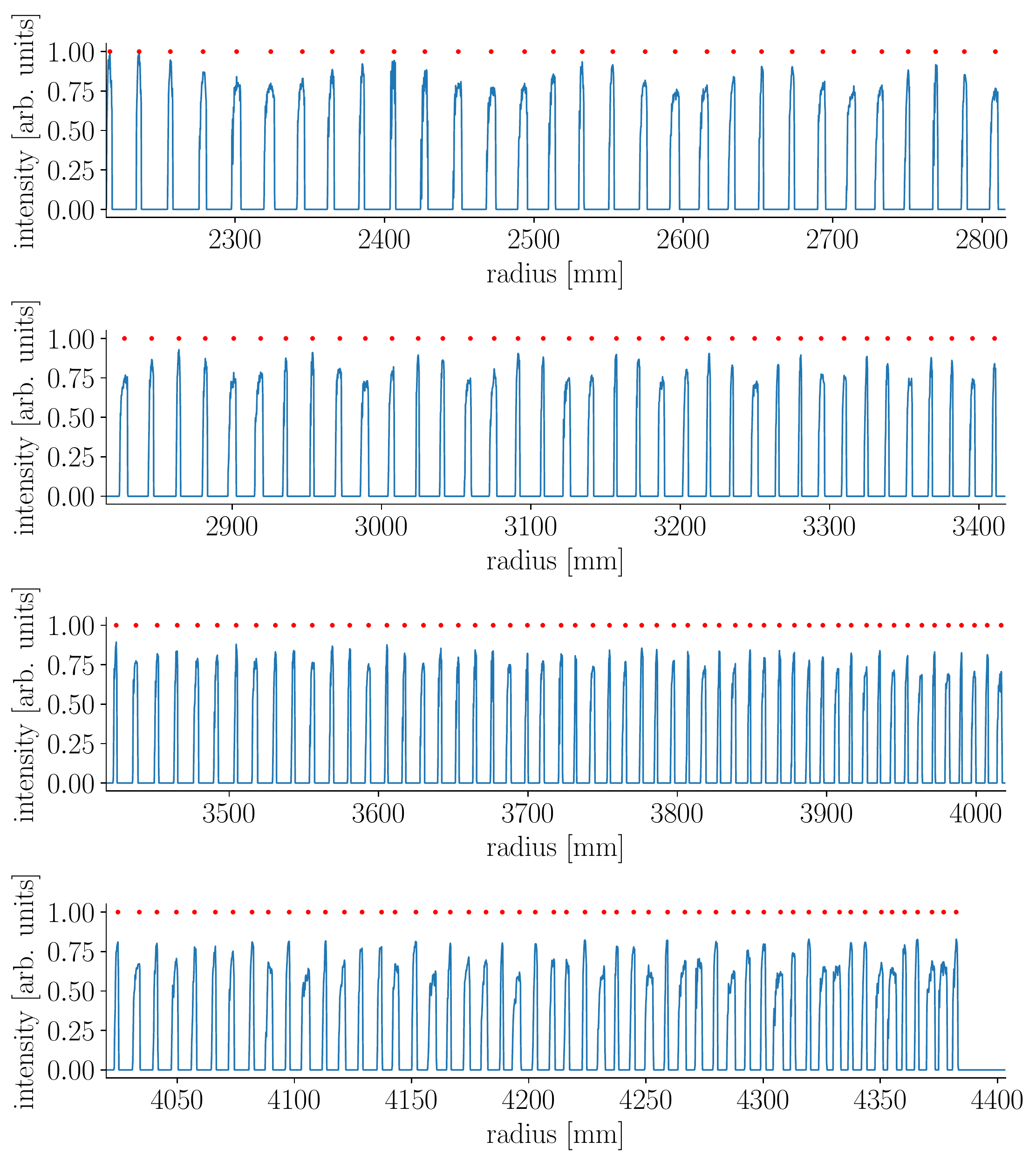}
    \caption{Histogram of the probe RRL measurement.\ The intensity is normalized.\
    The red dots mark detected peaks.}
    \label{fig:histogram_rrl1}
\end{figure}

To determine the radial beam position at each turn the radial profile from the
wire probe needs to be analyzed.\ This is done with a robust and
straightforward peak detection algorithm that searches for downward
zero-crossings in the smoothed first derivative with thresholds on minimum peak
value, area and slope.\ The identified peaks of the measurements are
indicated in \Figref{fig:histogram_rri2} and \Figref{fig:histogram_rrl1} with
red dots.

\begin{figure}[htp]
    \centering
    \includegraphics[width=1.0\columnwidth]{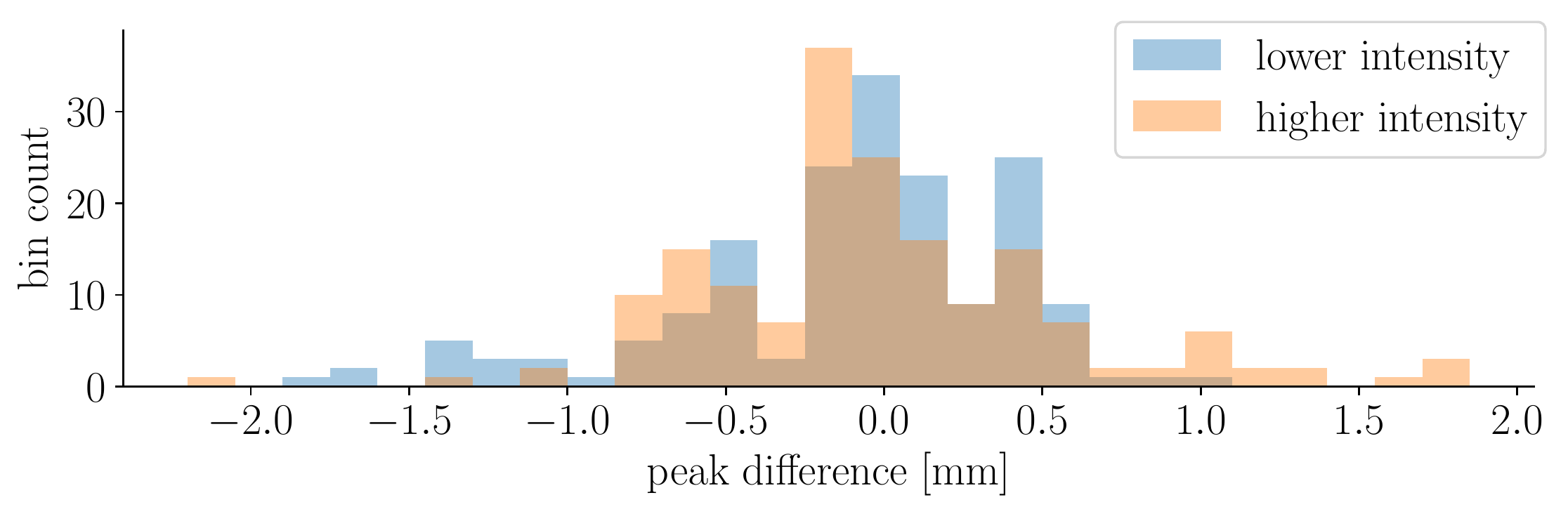}
    \caption{Histogram (bin width \SI[round-precision=2]{0.15}{mm}) of the changes
    in the peak positions for the RRL probe compared to the reference measurement
    of \Figref{fig:histogram_rrl1} for a lower (\SI{58}{\micro\ampere},  $\mu=\SI{-0.0913793103449}{mm}$,
    $\sigma=\SI{0.509659955183}{mm}$)
    and higher (\SI{108}{\micro\ampere}, $\mu=\SI{0.00344827586207}{mm}$, $\sigma=\SI{0.579624982974}{mm}$)
    intensity.\ The mean absolute error (MAE) taking both intensities is
    \SI{0.397988505747}{mm}.}
    \label{fig:peak_diffs53-55}
\end{figure}

In order to estimate the error of the measurements the
reference measurement
of \Figref{fig:histogram_rrl1} was compared to measurements with a lower
and higher beam current with the same machine condition.\ The histogram of the
changes in peak positions is shown in \Figref{fig:peak_diffs53-55}.\
As seen in the figure a change in the beam current does not influence the peak positions significantly.\

\subsection{Measurement of centered beam}
\label{sec:psi_cyc_measure_centered_beam}
The beam is extracted from the PSI Ring cyclotron using an electrostatic extractor,
the septum that is located in the gap between the last two turns.\
The standard production setup of the PSI Ring cyclotron makes use of a non-centered
beam such that the beam gap for the septum is enlarged by the beam precession.\ 
In order to obtain a proper scan of the turn pattern with a long radial probe, 
the beam has first to be centered accurately enough that individual turns are well 
separated.\ Only then, it is possible to accurately count the number of 
turns~\cite{Joho:CYC78, Stammbach:CYC92}.\ 

The beam centering of the PSI Ring cyclotron is determined by beam energy, radius 
and angle.\ The former is fixed by the extracted beam energy from Injector 2
but the latter can be manipulated by the last two injection magnets AND1 
and AND2 and the voltage of the electrostatic injection channel EIC (cf.\ \Figref{fig:psi_ring_cyclotron}).\ 
The centering of the beam is quantified by a numerical analysis of the data 
of a radial injection probe (RRI2).\ The radial positions $r_n$ of turn number 
$n$ can approximately be described by
\begin{equation*}
r_n=r_0+\left\langle{\frac{dr}{dn}}\right\rangle\,n+A\,\sin(2\,\pi\,\nu_r\,n+\phi)\,,
\end{equation*}
where $\nu_r$ is the radial tune, $A$ is the betatron amplitude and $\phi$ the 
betatron phase.\ The radius gain per turn $\left\langle{\frac{dr}{dn}}\right\rangle$
can be assumed to be approximately constant over a small range of turns where adjacent
turns do not overlap if $2A$ is smaller than the radius gain.\
For a centered beam the currents of AND1 and AND2 have to be chosen such that $A\approx 0$.\
A straightforward method, used also at PSI, is to measure the two-dimensional maps $A(I_1,I_2)$
and $\phi(I_1,I_2)$, where $I_1$ is the current in AND1 and $I_2$ the current 
in AND2 respectively.\ Then $A$ and $\phi$ can be interpolated.\

\begin{figure}[htp]
    \centering
    \includegraphics[width=\columnwidth]{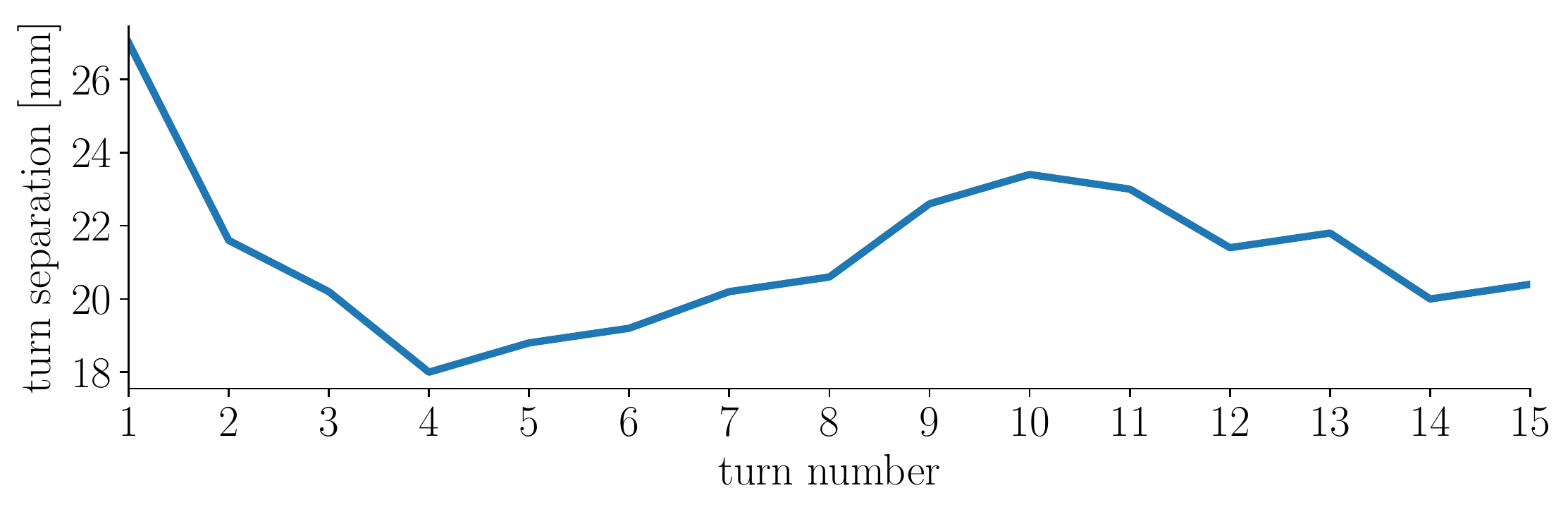}
    \caption{Turn separation among subsequent orbits measured at the probe RRI2
    (min.\ \SI{18.0}{mm} and max.\ \SI{27.0}{mm}).}
    \label{fig:turn_separation_rri2}
\end{figure}

\begin{figure}[htp]
    \centering
    \includegraphics[width=\columnwidth]{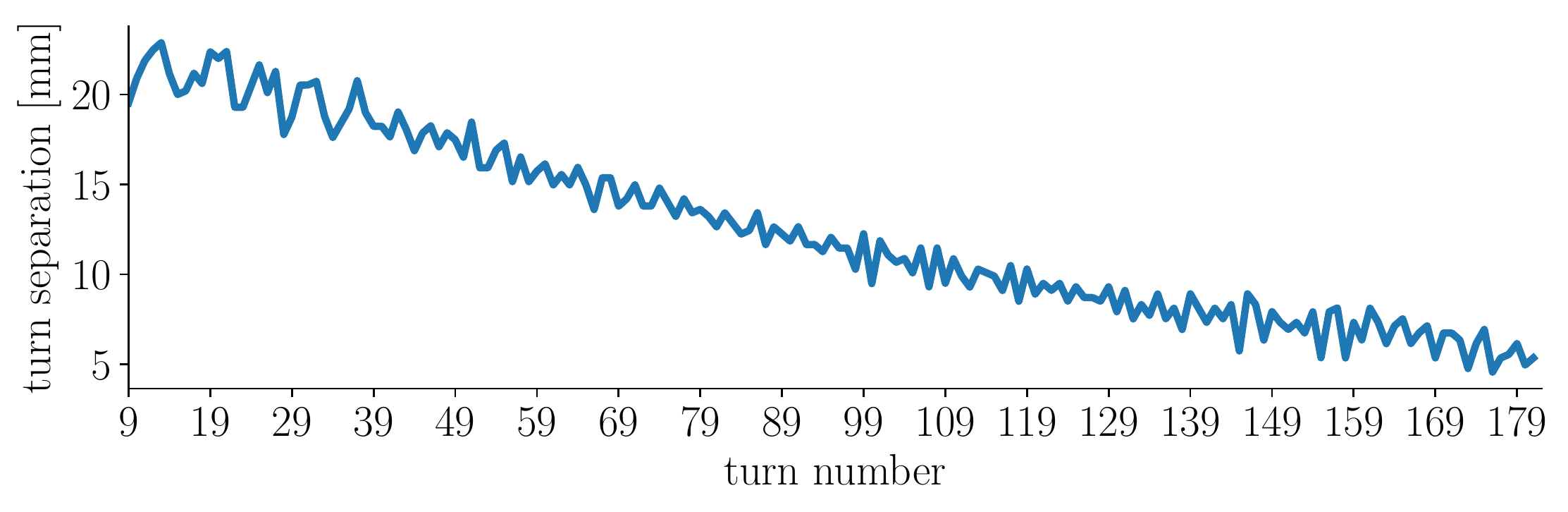}
    \caption{Turn separation among subsequent orbits measured at the probe RRL
    (min.\ \SI{4.56}{mm} and max.\ \SI{22.88}{mm}).}
    \label{fig:turn_separation_rrl1}
\end{figure}

The probe measurements used in this paper are performed with a beam intensity of
\SI{88}{\micro\ampere}.\ The beam profiles are given in \Figref{fig:histogram_rri2} and
\Figref{fig:histogram_rrl1}.\ The corresponding turn separation, i.e.\ the distance
between neighboring turns, at the probes is shown in \Figref{fig:turn_separation_rri2}
and \Figref{fig:turn_separation_rrl1}.\ At injection the turn separation is at maximum
\SI{27}{mm} which shrinks to \SI{6.09421052632}{mm} for the last \num{20}~turns.

\subsection{Trim coils}
\label{sec:psi_ring_cyclotron_trim_coils}
The PSI Ring cyclotron is equipped with 18 trim coils which allow compensation of
field errors and manipulation and optimization of the beam phase and isochronism.\
The trim coils are referred to as TC1 to TC18, from small to large radius.\ 
The trim coils TC6 to TC14 are only positioned on top of the odd numbered sector
magnets, i.e.\ SM1, SM3 etc.\ The other trim coils are on all magnets.\

The trim coils allow to change the field and to shift the beam phase.\ This can also
alter the energy gain per turn.\ 
Furthermore the trim coils enable, within certain limits, manipulation of the tune 
diagram of the ring cyclotron and thus to either avoid resonances or change 
their position and influence on the beam.\

As mentioned in the introduction, the lack of magnetic field data of the trim
coils makes it currently impossible to model the fields accurately.\ Therefore,
the simulation model developed in this paper is based on the average field profiles obtained
by measurements of the beam phase shifts as subsequently explained.

\subsubsection{Measurement fitting}
\label{sec:psi_ring_measurement_fitting}
\begin{figure}[htp]
        \centering
        \includegraphics[width=1.0\columnwidth]{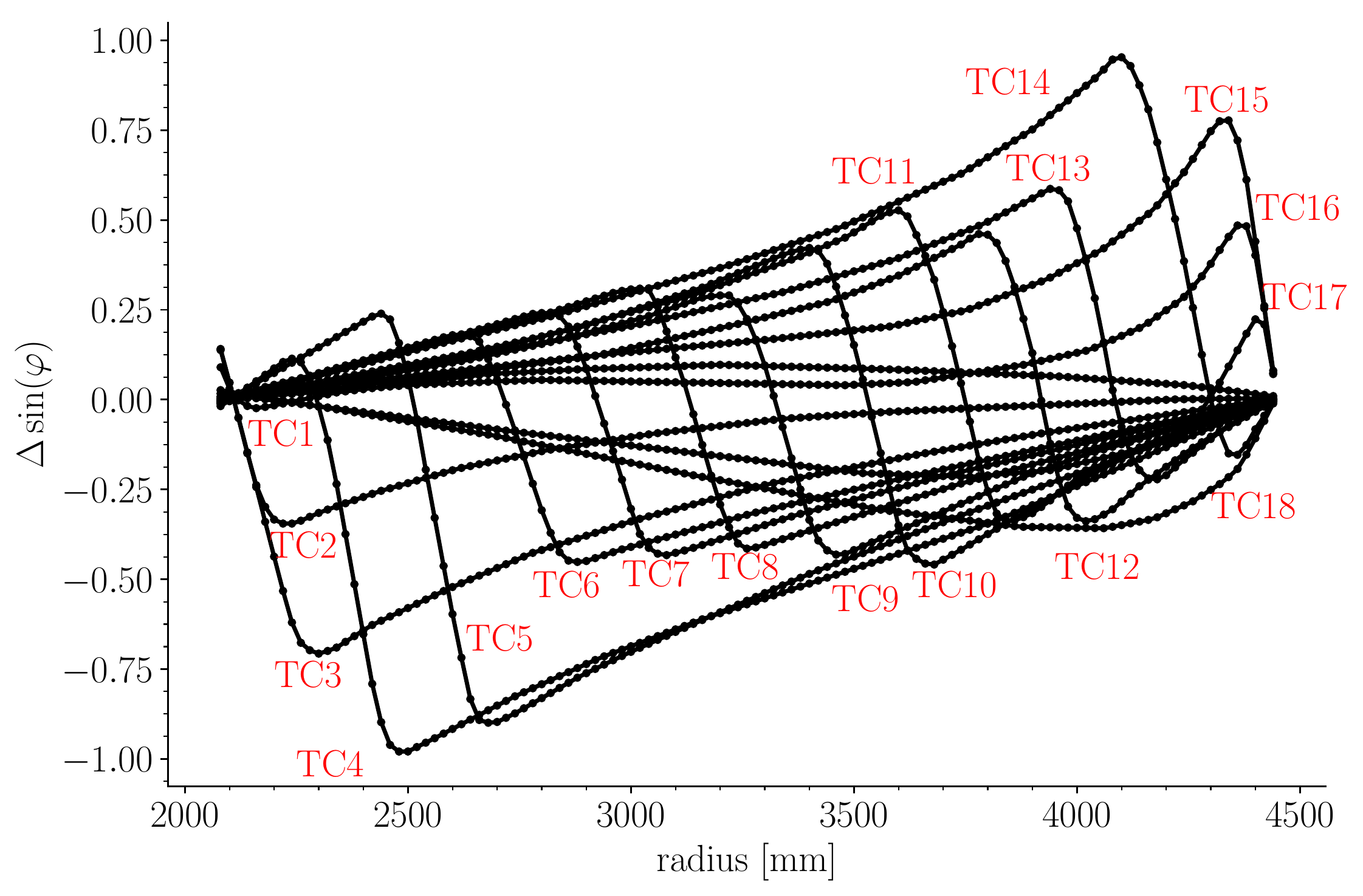}
        \caption{Measurement of the beam phase $\varphi$ shift due to trim coils~\cite{adam_stefan_1974_2556503}.}
        \label{fig:ring_phase}
\end{figure}

\begin{figure*}[htp]
    \centering
    \includegraphics[width=1.0\textwidth]{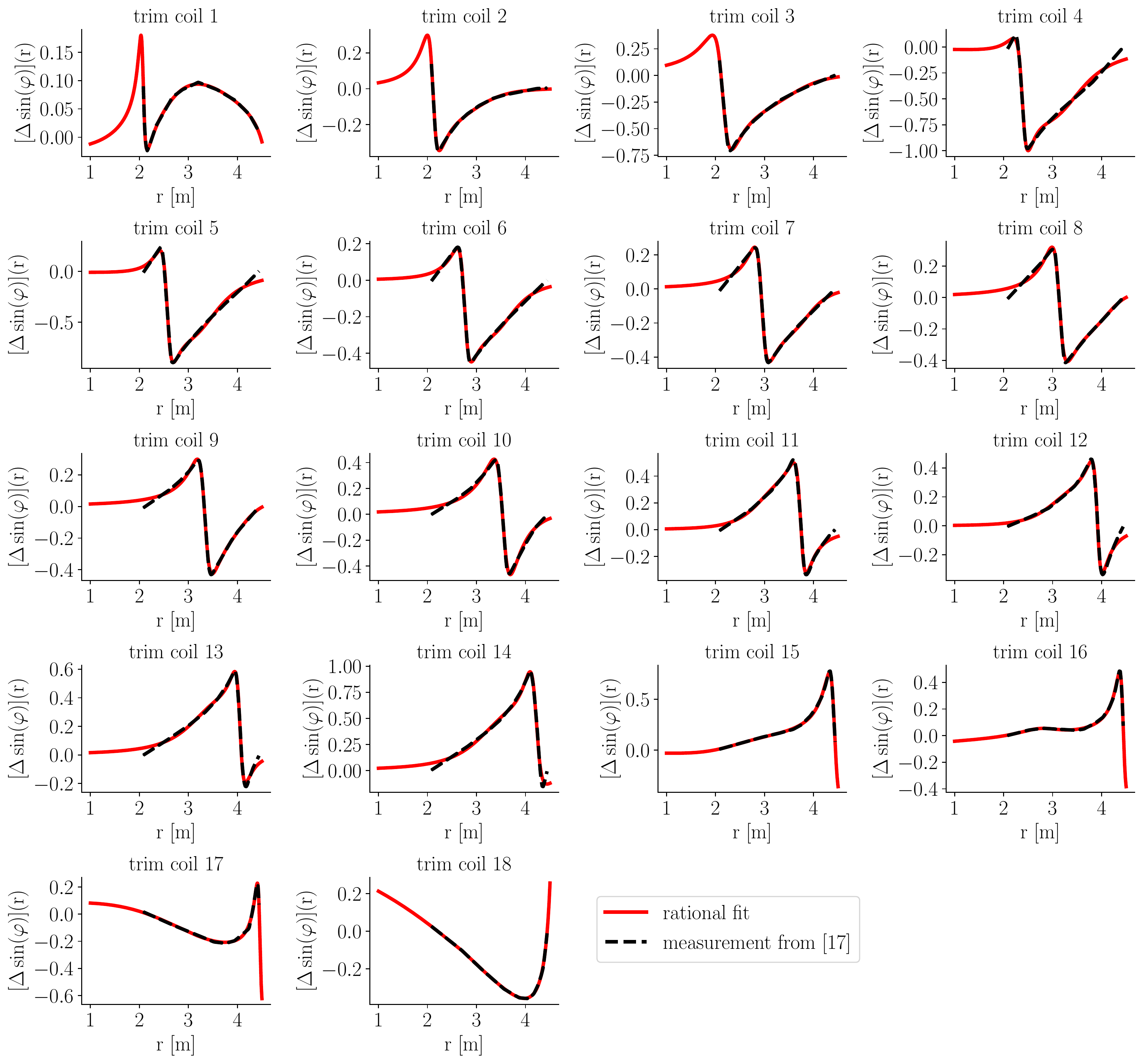}
    \caption{Fits of the $\Delta\sin(\varphi)$ measurements using rational functions.\ The change of the beam phase $\varphi$
    is induced by the trim coil fields.}
    \label{fig:fitted_phase}
\end{figure*}

\begin{figure*}[htp]
    \centering
    \includegraphics[width=1.0\textwidth]
    {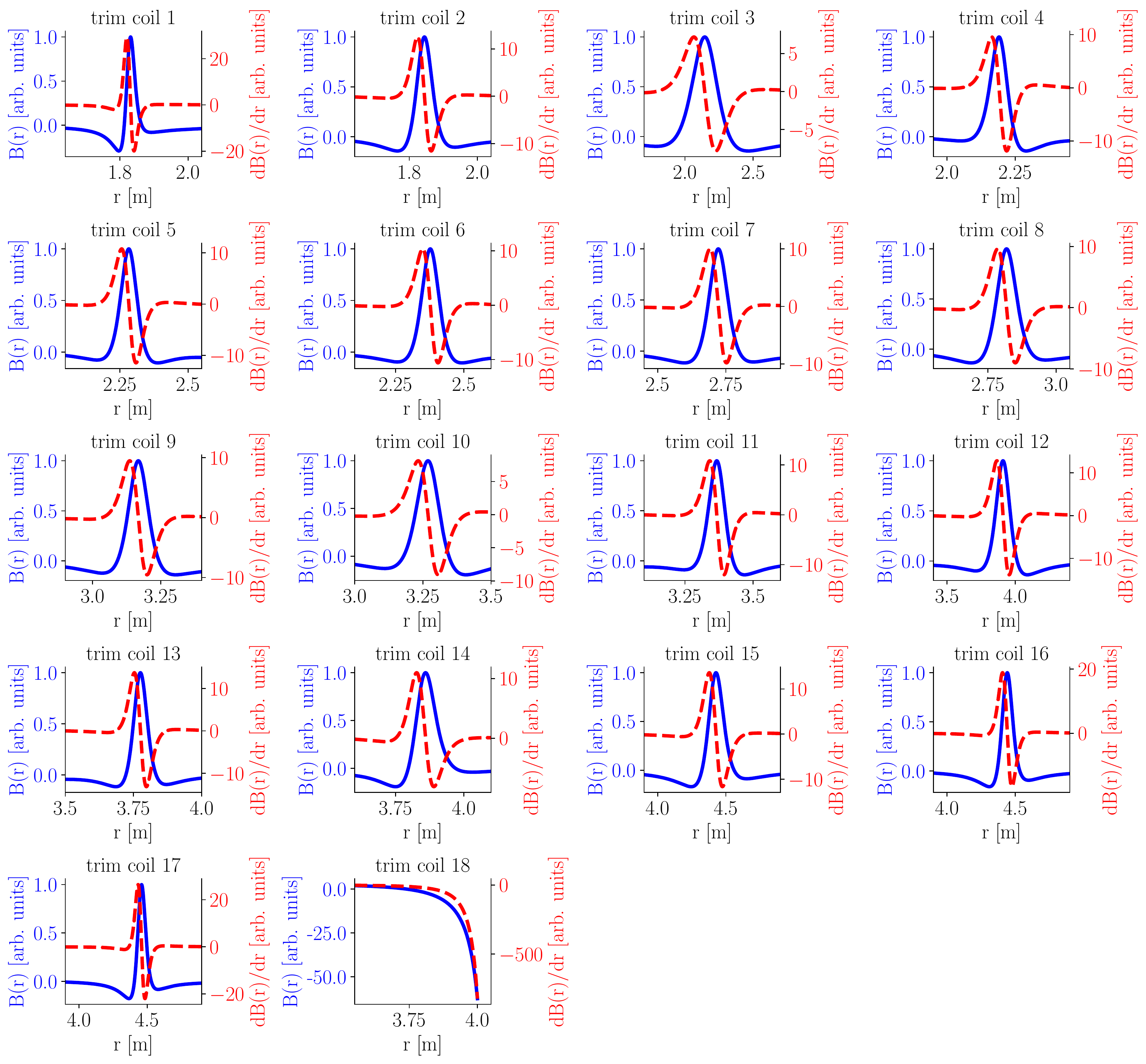}
    \caption{Magnetic fields and their radial derivative of the trim coils.}
    \label{fig:bfield_and_grad}
\end{figure*}

The presented trim coil model is based on measurements of $\Delta\sin(\varphi)$~\cite{adam_stefan_1974_2556503}
in \Eqref{eq:rational_fit} as depicted in \Figref{fig:ring_phase} with beam phase $\varphi$.\ As stated in
\cite{Parfenova:IPAC2016-TUPMR019}, the beam phase relates to $\Delta B_k$,
the magnetic field change due to trim coil $k$, by
\begin{equation}
    \Delta B_k \sim - \frac{qB(r)V(r)r}{E(r)\gamma (r)(\gamma (r)+1)} \frac{d\sin(\varphi)}{dr}
    \label{eq:Bk}
\end{equation}
with radius $r$, magnetic field $B(r)$, energy gain $V(r)$, charge $q$, kinetic energy $E(r)$ and
relativistic factor $\gamma (r)$.\ In the development of the trim coil model the simplified
relation
\begin{equation}
\Delta B_k \sim -  \frac{d\sin(\varphi)}{dr}
    \label{eq:Bk_nofactor}
\end{equation}
was used instead.\ Since the neglected factor of \Eqref{eq:Bk} varies little over the radial range of a
single trim coil, the negligence in \Eqref{eq:Bk_nofactor} doesn't deteriorate the model.\
In order to obtain the magnetic field magnitude as an additional degree of
freedom the numerical model relies on normalized fields as discussed later.\ 
Each trim coil phase data was approximated by a rational function, i.e.\
\begin{equation}
    [\Delta\sin(\varphi)](r) \approx \frac{f(r)}{g(r)}
                           = \frac{\sum_{i=0}^{n}a_ir^i}{\sum_{j=0}^{m}b_jr^{j}}
    \label{eq:rational_fit}
\end{equation}
with $m, n\in \mathbb{N}_0$ and $m>n$.\ The coefficients were computed by a
\python{} script using the non-linear least-squares method of \scipy{} \cite{Scipy}
where $n = 2$, $m=4$ for trim coils TC2 - TC15 and $n = 4$, $m=5$
for TC1 and TC16 - TC18 respectively.\ The fits of the data are shown in
\Figref{fig:fitted_phase}.\ The selection of
the parameters $n$ and $m$ was done empirically trying to keep the polynomial degree
small.\ As a result of \Eqref{eq:rational_fit} the corresponding magnetic field is
therefore simply given by
\begin{equation}
B(r) \sim \frac{d}{dr} \frac{f(r)}{g(r)} = \frac{f(r)g'(r) - f'(r)g(r)}{g^2(r)},
\label{eq:bfield}
\end{equation}
with $f'(r) \equiv df(r)/dr$.\ The normalized magnetic field and its derivative of each trim coil
are depicted in \Figref{fig:bfield_and_grad}.\

    \lstset{
    language=C++,
    morekeywords={*,TRIMCOIL,
                  TYPE,
                  RMIN,
                  RMAX,
                  BMAX,
                  COEFNUM,
                  COEFDENOM,
                  CYCLOTRON,
                  TRIMCOILTHRESHOLD,
                  PROBE,
                  XSTART,
                  XEND,
                  YSTART,
                  YEND
                  },
    keywordstyle=\color{blue},
    commentstyle=\color{green!50!black},
    stringstyle=\color{red},
    frame=tb,
    framerule=1pt,
    captionpos=b,
    basicstyle=\footnotesize\ttfamily
}

\section{OPAL}
The open source library \opal{} \cite{opal:1} is a parallel electrostatic
Particle-In-Cell (ES-PIC) framework for large-scale particle accelerator
simulations.\ In the sequel the \opal-cycl flavor that is used for all simulations
in this study is also referred as \opal.\ The particles are evolved in time $t$ by either a fourth
order Runge-Kutta or a second order Leapfrog according to the collisionless Boltzmann
(or Vlasov-Poisson) equation
\begin{equation*}
    \frac{df}{dt} = \frac{\partial f}{\partial t}
                  + \vvec\cdot\nabla_\xvec f
                  + \frac{q}{m_0}\left(\Evec +
                                       \vvec \times \Bvec\right)
                    \cdot \nabla_\vvec f
                  = 0,
\end{equation*}
with charge $q$, mass $m_0$ and the six-dimensional particle density function
$f(\xvec, \vvec, t)$ with $(\xvec, \vvec)\in\mathbb{R}^{3\times 3}$.\ The electromagnetic
fields $\Evec\equiv\Evec(\xvec, t)$ and $\Bvec\equiv\Bvec(\xvec, t)$ consist of a
bunch internal and
external contribution.\ The bunch self-field is obtained in the beam rest frame by
either a FFT Poisson solver or a Smoothed Aggregation Algebraic Multigrid (SAAMG) \cite{ADELMANN20104554}
solver that is able to handle arbitrary accelerator geometries.\

The following subsections highlight three features of \opal{} that were used as
well as extended for the purpose of this study.

\subsection{Probe element and peak detection}
In the cyclotron flavor of \opal{} the probe is a special element placed on the midplane in
Cartesian coordinates to record particles in simulations.\ The origin is the
machine center, therefore, the positioning according to \Eqref{eq:probe_positioning}
with \Figref{fig:probe_positioning} is directly applicable.\ The \opal{} syntax is
shown in \Figref{fig:probe_command}.\ 
Since \opal{} has fixed time steps, 
particle position recording at the probe is done by a linear extrapolation of the particle direction
from the closest tracking point towards the probe axis.
In order to compare measurement and simulation
the original description in \opal{} was extended to write a particle histogram
and a file collecting the peak locations.\
In single particle tracking the peak and thus turn detection is
trivial.\ The localization of a turn in a multi particle simulation is achieved
by summing up all radii where particles hit the probe.\ The mean is then
determined as the orbit radius.
\begin{figure}[htp]
    \centering
    \begin{minipage}{1.0\columnwidth}
        \begin{lstlisting}
probe: PROBE, XSTART = ..., // [mm]
              XEND   = ..., // [mm]
              YSTART = ..., // [mm]
              YEND   = ..., // [mm]
              ...;
        \end{lstlisting}
    \end{minipage}
    \caption{\opal{} input command for probe elements.}
    \label{fig:probe_command}
\end{figure}

\subsection{Multi-objective optimization}
Since release version 2.0.0, \opal{} is equipped with a multi-objective genetic
algorithm \nsga{} (non-dominated sorting genetic algorithm) implementation \cite{20.500.11850/72925}.\
The new \opal{} feature was already applied in \cite{Neveu:2013ues}.\ Initially a
population of $n$ randomly spawned individuals within a predefined hyperspace of
design variables is created after which a new set of individuals for the next
population within the same bounds is selected by mixing the current $k$-fittest
individuals based on a crossover pattern
as well as a gene mutation algorithm.\ The fitness of
an individual is determined by the evaluation of user-defined functions on
the objectives that in this paper are the peak differences between measurement and
simulation.\ In this regard the framework was extended to be able to read peak
files and to evaluate the $l_{\infty}$-error and $l_{2}$-error norms on the parsed peak
data.\ Furthermore, an expression to check the number of turns in simulations was
added that served as an individual constraint.\ One of the main benefits of MOGA is
its ability to search in parallel as it can evaluate all individuals of a single generation at the same time.

\subsection{New trim coil model}
The existing trim coil model in \opal{} \cite{PhysRevSTAB.14.054402}
was especially designed to fit the shape of
TC15 of the PSI Ring cyclotron because its interest focused
on the turns close to extraction.\ Furthermore, the field was contributed
not only local to the sector magnets but continuous smeared out on \SI{360}{\degree}.\
The new model uses a more general description by rational functions as described
in \Secref{sec:psi_ring_measurement_fitting}.\ This representation of the field
allows a simple analytical differentiation to obtain the necessary derivative
for the magnetic field interpolation to the position of each particle.\ That way
the model is not restricted to the specific shape of TC15 in \cite{PhysRevSTAB.14.054402}.\

The new trim coil model does not support an azimuthally limited field definition.\ However, the
trim coil fields are restricted to the sector magnets by a user-defined threshold
that is the lower limit to apply the additional fields.\ The implementation of the
trim coils assumes normalized polynomial coefficients such that the maximum value
of the field is $1.0$, thus, the maximum field strength $B_{\max}$ is
an auxiliary tuning parameter, i.e.\
\begin{equation*}
    \mbox{TC}(r) = B_{\max}\frac{\sum_{i=0}^{n}a_ir^i}{\sum_{j=0}^{m}b_jr^{j}}
\end{equation*}
with $n,m\in\mathbb{N}_{0}\wedge r\in\left[r_{\min}, r_{\max}\right]$.\
Secondly, the trim coil field is restricted in the radial direction
by two extra parameters $r_{min}$ and $r_{max}$ to allow more flexibility.\
Nevertheless, the bounds have to be selected carefully to avoid a discontinuity
in the magnetic field.\ In azimuthal direction the
implementation uses a linear decaying field to prevent the previously mentioned
issue.\
Since the functions $f(r)$
and $g(r)$ in \Eqref{eq:rational_fit} are polynomials in radius $r$, the derivative
\Eqref{eq:bfield} is a rational function again.\ The model can therefore
accept either the phase or magnetic field as input.

A template of a trim coil definition in an \opal{} input file is given in
\Figref{fig:tc_command}.\ The parameter \textsf{TYPE} specifies if the polynomial
represents the phase \textsf{PSI-PHASE} or the magnetic
field \textsf{PSI-BFIELD}.\
In order to be applied the trim coil elements have to be appended to a list in
the cyclotron command as depicted in \Figref{fig:cycl_command}.

\begin{figure}[htp]
    \centering
    \begin{minipage}{1.0\columnwidth}
        \begin{lstlisting}
tc1: TRIMCOIL, TYPE = "PSI-PHASE",
               RMIN = ..., // inner radius [mm]
               RMAX = ..., // outer radius [mm]
               BMAX = ..., // B-field peak [T]
               COEFNUM   = {a0, a1, a2, ...},
               COEFDENOM = {b0, b1, b2, ...};
        \end{lstlisting}
    \end{minipage}
    \caption{\opal{} input command for trim coils.}
    \label{fig:tc_command}
\end{figure}

\begin{figure}[htp]
    \begin{minipage}{1.0\columnwidth}
        \begin{lstlisting}
                 // lower limit of TC input [T]
Ring: CYCLOTRON, TRIMCOILTHRESHOLD = ...,
                 TRIMCOIL = {tc1, tc2, tc3, ...}
                 ...
                 ;
        \end{lstlisting}
    \end{minipage}
    \caption{Additional trim coil input arguments for the cyclotron element
    definition in \opal.}
    \label{fig:cycl_command}
\end{figure}

    \section{Turn Pattern Matching}
\label{sec:turn_pattern_matching}
A measure for the quality of the pattern matching is 
the maximal peak difference between measurement $m$ and simulation $s$, i.e.\
\begin{equation}
\min \max_{i=1\dots N} \left|r_{i}^{m} - r_{i}^{s}\right|,
\label{eq:matching_quality}
\end{equation}
where $N$ is the number of turns and $r_{i}^{m}$ and $r_{i}^{s}$ are the $i$-th turn radii.\

An iterative process to get a model that is in good agreement with measurements 
applied multi-objective optimization and 
local search.\ Furthermore, the input parameter space 
between different optimizations was flexible, i.e.\ design 
variables (DVARs) were added and removed.\ The selection of the 
design variables is described in detail in the subsequent section.\

\begin{figure}[htp]
    \includegraphics[width=1.0\columnwidth]{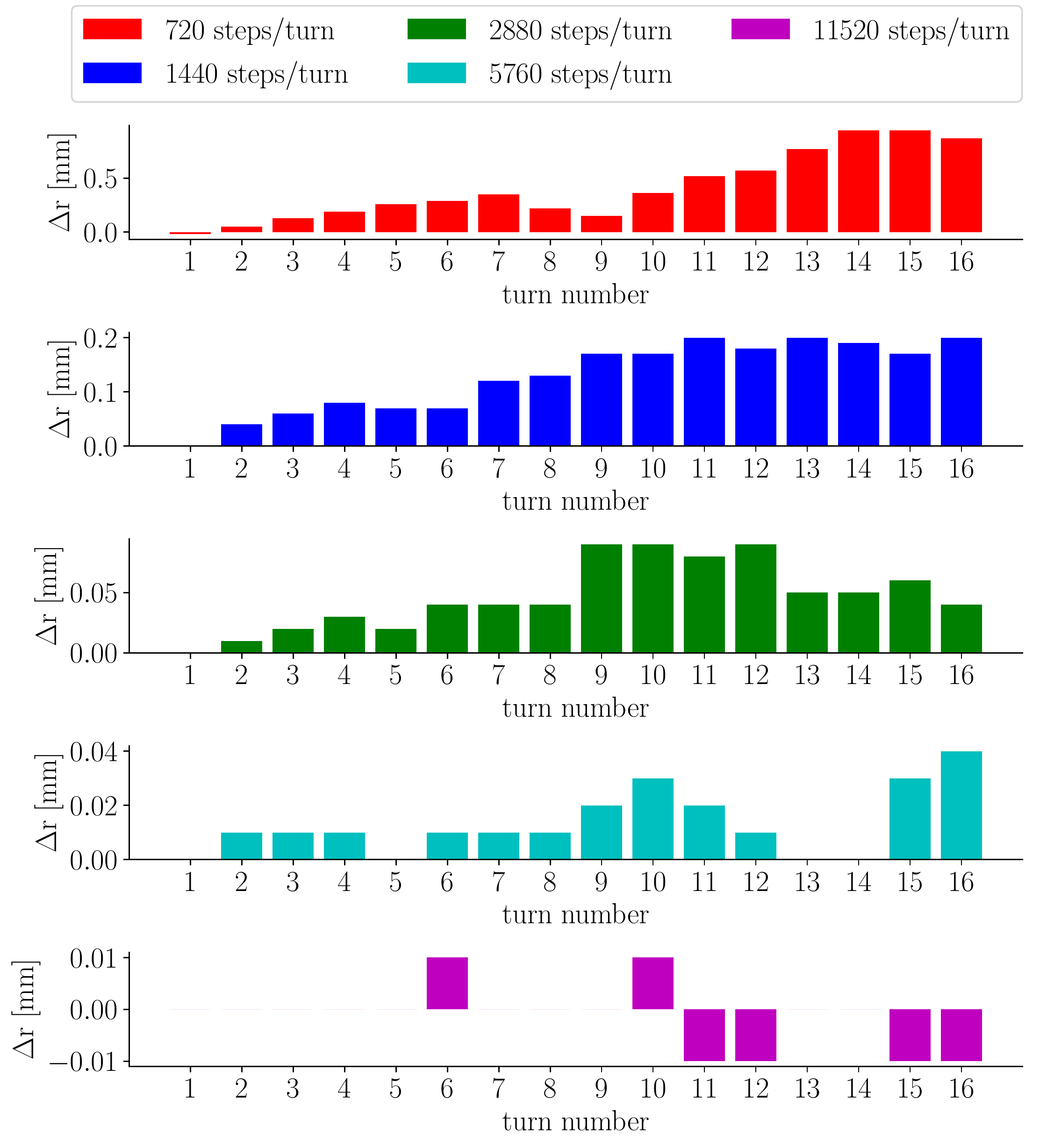}
    \caption{Difference of the turn radius $\Delta r$ at probe RRI2 between simulations due to
    the number of integration steps per turn.\ 
    The reference simulation uses \num{23040} steps/turn.\ 
    The input parameters of the simulation
    are optimized for \num{2880} steps/turn.}
    \label{fig:RRI2_radius_error_simulation}
\end{figure}

\begin{figure}[htp]
    \includegraphics[width=1.0\columnwidth]{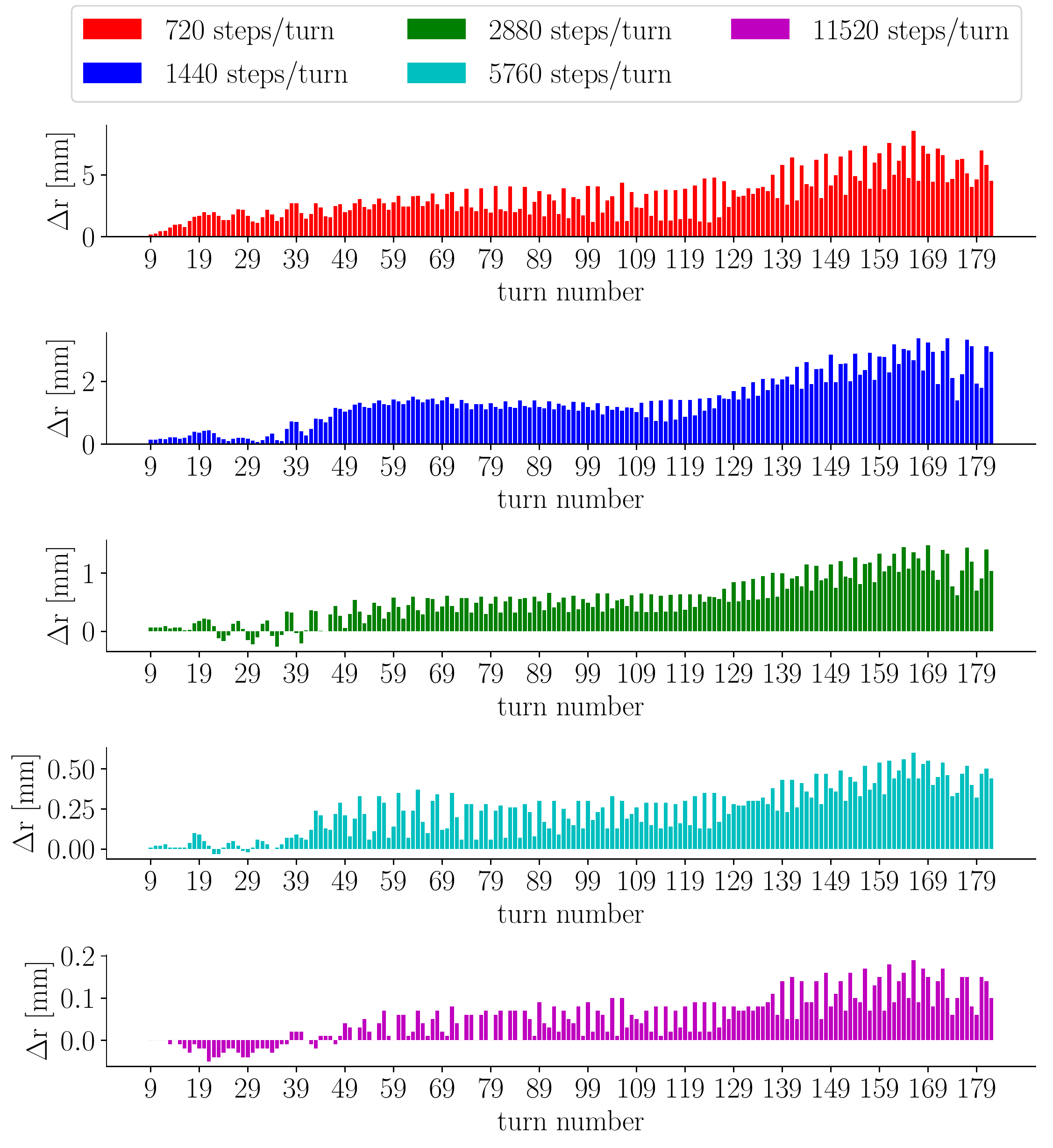}
    \caption{Difference of the turn radius $\Delta r$ at probe RRL between simulations due to
    the number of integration steps per turn.\ 
    The reference simulation uses \num{23040} steps/turn.\ 
    The input parameters of the simulation
    are optimized for \num{2880} steps/turn.}
    \label{fig:RRL1_radius_error_simulation}
\end{figure}

Due to the decrease of the turn separation as discussed in
\Secref{sec:psi_cyc_measure_centered_beam} and the increase of the circumference
of the machine the choice of the number of steps per turn in simulation needs to be chosen
carefully to obtain reliable results.\ Furthermore, the extrapolation method that
is used to get the point where the particle hits the probe depends also on this time
discretization.\ After a comparison between different number of steps per turn and a reference simulation
with \num{23040} integration steps per turn
(cf.\ \Figref{fig:RRI2_radius_error_simulation} and \Figref{fig:RRL1_radius_error_simulation}),
the optimal number of steps per turn for the
fourth order Runge-Kutta integrator w.r.t.\ accuracy and runtime turned out to
be \num{2880}.\ It compares to the reference simulation with a maximum absolute
error of \SI{1.48}{mm}, mean absolute error (MAE) \SI{0.560344827586}{mm} and mean squared error
(MSE) \SI{0.449387356322}{mm\squared}.\ 
The reference simulation is selected based on the observation
that the turn radii differ only in the order of $\mathcal{O}(\num{0.19})\ \si{mm}$ at
the probes compared to \num{11520} steps per turn.\ A significant improvement is
only achieved with an integrator of higher order.\ 

\subsection{Design variable selection}
As previously mentioned the selection of the design variables 
wasn't obvious at first.\ Initially, the beam 
injection parameters and the peak magnetic field of the first trim coils were 
considered in order to match the first turns at injection.\ However, the idea of 
matching basically turn by turn starting at injection failed soon since the turn
pattern difference between measurement and simulation started to diverge at later
turns due to the wrong energy gain per turn.\ As a consequence the RF cavity 
parameters together with a constraint on the number of completed turns were added to
guide the optimization towards solutions with the right number of turns.\ 
For RF cavities their voltages and positions were varied where a position encompasses
the angle, radial position and displacement from the global center.\
For the flat top cavity also the phase angle was added as a parameter.\
The angle between RRI2 and RRL and their radial position were varied in 
order to smooth the transition between the probes.\

%

Since TC18 is turned off, it was not added as design variable.\ Also TC17 that
influences the last few turns was discarded since the extraction channel is not
simulated.\ However, these turns are still corrected in simulation by TC16.\

\begin{figure}[htp]
    \centering
    \tikzset{
        parallel segment/.style={
        segment distance/.store in=\segDistance,
        segment pos/.store in=\segPos,
        segment length/.store in=\segLength,
        to path={
        ($(\tikztostart)!\segPos!(\tikztotarget)!\segLength/2!(\tikztostart)!\segDistance!90:(\tikztotarget)$) -- 
        ($(\tikztostart)!\segPos!(\tikztotarget)!\segLength/2!(\tikztotarget)!\segDistance!-90:(\tikztostart)$)  \tikztonodes
        }, 
        segment pos=0.5,
        segment length=10ex,
        segment distance=3mm,
        },
    }
    \begin{tikzpicture}[scale=0.75]
        \node at (0, 0) {\includegraphics[width=\columnwidth]{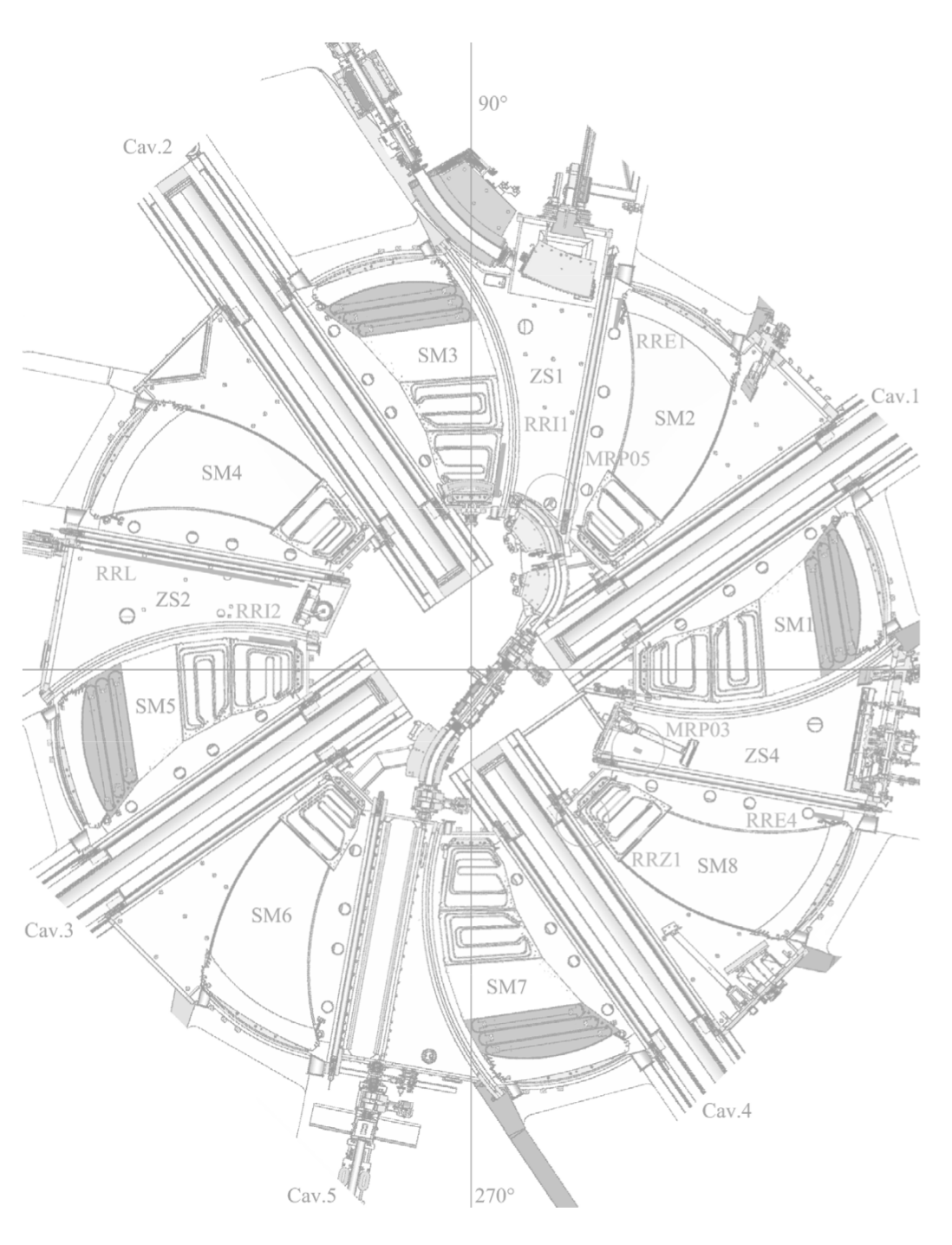}};
        
        \draw[line width=2.4pt, -stealth, rotate around={-30:(4.5, 2)}] (4.5, 2) arc (50:100:1.4)
            node[left] {\bf\circled{5}};
        
        \draw[line width=2.4pt, -stealth, rotate around={60:(-3, 4.5)}] (-3, 4.5) arc (50:100:1.4)
            node[left] {\bf\circled{5}};
        
        \draw[line width=2.4pt, -stealth, rotate around={160:(-4.7, -3.3)}] (-4.7, -3.3) arc (50:100:1.4)
            node[below] {\bf\circled{5}};
        
        \draw[line width=2.4pt, -stealth, rotate around={250:(2.8, -5.4)}] (2.8, -5.4) arc (50:100:1.4)
            node[right] {\bf\circled{5}};
        
        \draw[line width=2.4pt, blue, |-|, rotate around={35:(0.95, -0.48)}] (0.95, -0.48)  -- (5.95, -0.48);
        
        \draw[line width=2.4pt, stealth-stealth, rotate around={35:(0.95, -0.48)}]
            (0.95, -0.48) to[parallel segment] (5.95, -0.48) node[above] {\bf\circled{1}};
        
        \draw[line width=2.4pt, blue, |-|, rotate around={-55:(-0.2, 0.45)}] (-0.2, 0.45)  -- (-6.4, 0.45);
        
        \draw[line width=2.4pt, stealth-stealth, rotate around={-55:(-0.2, 0.45)}]
            (-0.2, 0.45) to[parallel segment] (-6.4, 0.45) node[left] {\bf\circled{1}};
        
        \draw[line width=2.4pt, blue, |-|, rotate around={35:(-1.05, -0.75)}] (-1.05, -0.75)  -- (-6.05, -0.75);
        
        \draw[line width=2.4pt, stealth-stealth, rotate around={35:(-1.05, -0.75)}]
            (-1.05, -0.75) to[parallel segment] (-6.05, -0.75) node[below] {\bf\circled{1}};
        
        \draw[line width=2.4pt, blue, |-|, rotate around={-55:(0.1, -1.6)}] (0.1, -1.6)  -- (5.1, -1.6);
        
        \draw[line width=2.4pt, stealth-stealth, rotate around={-55:(0.1, -1.6)}]
            (0.1, -1.6) to[parallel segment] (5.1, -1.6) node[right] {\bf\circled{1}};
        
        \draw[line width=1.2pt, black, dashed] (-0.06, -6) -- (-0.06, 6.0);
        \draw[line width=1.2pt, black, dashed] (-6, -0.58) -- (6.0, -0.58);
        \draw[line width=1.6pt, black, dashed, rotate around={36:(0.95, -0.55)}] (0.1, 0) -- (1.0, 0);
        
        \draw[line width=1.6pt, red, -stealth, rotate around={0:(0.65, 0)}] (0.65, -0.05) -- (0.95, -0.48)
            node[below] {\bf\circled{2}};
        
        \draw[line width=2.4pt, -stealth, rotate around={200:(-1.35, -5)}] (-1.35, -5) arc (50:100:1.3)
            node[below=0.18cm] {\bf\circled{8}};
        
        \draw[line width=2.4pt, purple, |-|, rotate around={87:(-0.88, -2.4)}] (-0.88, -2.4)  -- (-4, -2)
            node[below] {\textcolor{black}{\bf\circled{\textcolor{black}{10}}}};
        
        \draw[line width=2.4pt, stealth-stealth, rotate around={87:(-0.88, -2.4)}]
            (-0.88, -2.4) to[parallel segment] (-4, -2) node[above right] {\bf\circled{3}};
        
        \draw[line width=1.6pt, black, dashed, rotate around={-100:(-0.06, -0.58)}] (-0.06, -0.58) -- (1.9, -0.58);
        \draw[line width=1.6pt, teal, stealth-, rotate around={0:(-0.88, -2.4)}] (-0.88, -2.4) -- (-0.4, -2.5)
            node[below] {\bf\circled{4}};
        
        \node[magenta] at (0.4, 1.5) {\bf\circled{6}};
        
        \draw[line width=2.4pt, green!40!black] (-2.2, 0.43) -- (-4.8, 0.9)
            node[above right] {\bf RRL};
        \draw[line width=2.4pt, green!40!black] (-2.2, -0.24) -- (-2.8, -0.15)
            node[above left] {\bf RRI2};
        \node[green!40!black] at (-2.2, 0.1) {\bf\circled{7}};
        
        \node at (3.8, 0.1)   {\bf\circled{9}};
        \node at (2.4, 2.5)   {\bf\circled{9}};
        \node at (-0.4, 3.2)  {\bf\circled{9}};
        \node at (-3.2, 1.9)  {\bf\circled{9}};
        \node at (-3.9, -1.0) {\bf\circled{9}};
        \node at (-2.5, -3.6) {\bf\circled{9}};
        \node at (0.4, -4.4)  {\bf\circled{9}};
        \node at (3.0, -3.0)  {\bf\circled{9}};
    \end{tikzpicture}
    \caption{Design variables in context of the PSI Ring cyclotron.\ In appendix
    \Tabref{tab:dvars} is a description of each variable.\
    Legend:
    \ding{172} rmainshift1 - rmainshift4, vmaincav1 - vmaincav4;
    \ding{173} pdismain1 - pdismain4;
    \ding{174} rftshift;
    \ding{175} pdisft;
    \ding{176} phimain1 - phimain4;
    \ding{177} benergy, prinit, phiinit, rinit;
    \ding{178} rrla, rrlphi, rrlshift, rri2a, rri2phi, rri2shift;
    \ding{179} phift;
    \ding{180} tc01mb - tc16mb;
    \ding{181} phirfft, vftcav.}
    \label{fig:psi_ring_cyclotron_dvars}
\end{figure}

The final list of \num{48} DVARs is given in \Tabref{tab:dvars} of the 
appendix.\ As a further clarification they are also depicted in the drawing of 
the cyclotron in \Figref{fig:psi_ring_cyclotron_dvars}.\ The angle between the 
probes RRI2 and RRL as indicated by \ding{178} in the plot is adjusted using 
the variables $a$, $\varphi$ of \Eqref{eq:probe_positioning} while  keeping the 
length of each probe $s\in\left[s_0+t, s_1+t\right]$ with $t\in\mathbb{R}$ 
fixed.\

\subsection{Model simplifications}
Beside numerical approximations on the design of the new trim coil model the 
large number of DVARs (i.e.\ \num{48}) and objectives (i.e.\ \num{182} turns) 
required further simplifications on the optimization approach as explained in 
the following.\

\subsubsection{Aggregation of turns}
\label{sec:aggregation_of_turns}
In case of the PSI Ring cyclotron the number of objectives (i.e.\ turns) is \num{182}.\
In order to reduce this space multiple turns were clustered to single objectives
$\sigma_{[l,u]}$ with turns in the range $[l, u]\in [1, 182]$ by either
the $l_2$-error norm
\begin{equation*}
    \sigma_{[l, u]} = \frac{1}{N}\sqrt{\sum_{i=l}^{u} (r_{i}^{m} - r_{i}^{s})^{2}}
\end{equation*}
with $N=u - l + 1$ the number of aggregated turns or the $l_{\infty}$-error norm
\begin{equation*}
    \sigma_{[l,u]} = \max_{i=l\ldots u} \left|r_{i}^{m} - r_{i}^{s}\right|
\end{equation*}
where $r_{i}^{m}$ and $r_{i}^{s}$ are the $i$-th turn radii of the measurement
and simulation respectively.\ The $l_\infty$-norm suits our definition
of the measure for the pattern matching quality, i.e.\ \Eqref{eq:matching_quality}.\

\subsubsection{Reduction of trim coil support}
Due to the field overlap of neighboring trim coils a valid assumption is the
partial cancellation of the field tails.\ That's why the model uses a reduced radial
support.\ Only trim coil TC1 uses the full range on
the lower half.\

\subsubsection{Location of trim coils}
In order to limit the trim coil field in the azimuthal direction the user 
provides a lower bound of the magnetic field by the attribute \textsf{TRIMCOILTHRESHOLD}
(cf.\ \Figref{fig:cycl_command}) above which the trim coil field is applied.\
This is a limitation of the new model since the real machine provides the field of all 
trim coils only on specific sector magnets (cf.\ \Secref{sec:psi_ring_cyclotron_trim_coils}).\

\subsubsection{Single particle tracking}
The radial profiles are measured using a low intensity beam, i.e.\ 
\SI{88}{\micro\ampere}.\
The negligence of space charge in order to lower the time to solution of a
single simulation is therefore a reasonable assumption.\ A further 
simplification to
single particle tracking is motivated by the observation that peaks are
detected at the centroid of the beam (cf.\ \Figref{fig:histogram_rri2} and
\Figref{fig:histogram_rrl1}).\ This reduces the time to model the full
machine to approximately \SI{2}{\second} on a single core.\

\subsection{Multi-objective optimization}
The dimension of the design variable space required a rather large
number of individuals per generation in order to sample the space sufficiently.\
All optimizations were performed on Piz Daint \cite{PizDaint}, a supercomputer of the
Swiss National Supercomputing Centre (CSCS).\
Due to its hardware architecture where a node is equipped with \num{2}
Intel Xeon E5-2695 v4 $@$ \SI{2.10}{GHz} ($2 \times 18$ cores, \SI{64/128}{GB} RAM)
processors, a total number of \num{8062} individuals was selected in which two
cores of the \num{224} nodes ($\num{36}\cdot\num{224} - 2 = 8062$) were reserved
for individual post-processing and bookkeeping.\
Since a single objective according to \Secref{sec:aggregation_of_turns} didn't perform well,
the turns were grouped into a total of finally six objectives.\ A reason might
be local optima from which MOGAs are more likely to escape as discussed in \cite{1688438}.\
While RRI2 was kept as a single objective, RRL was split into five objectives where
each had approximately the same amount of turns and was influenced by a single trim coil only.\

\begin{figure}[htp]
    \centering
    \includegraphics[width=\columnwidth]{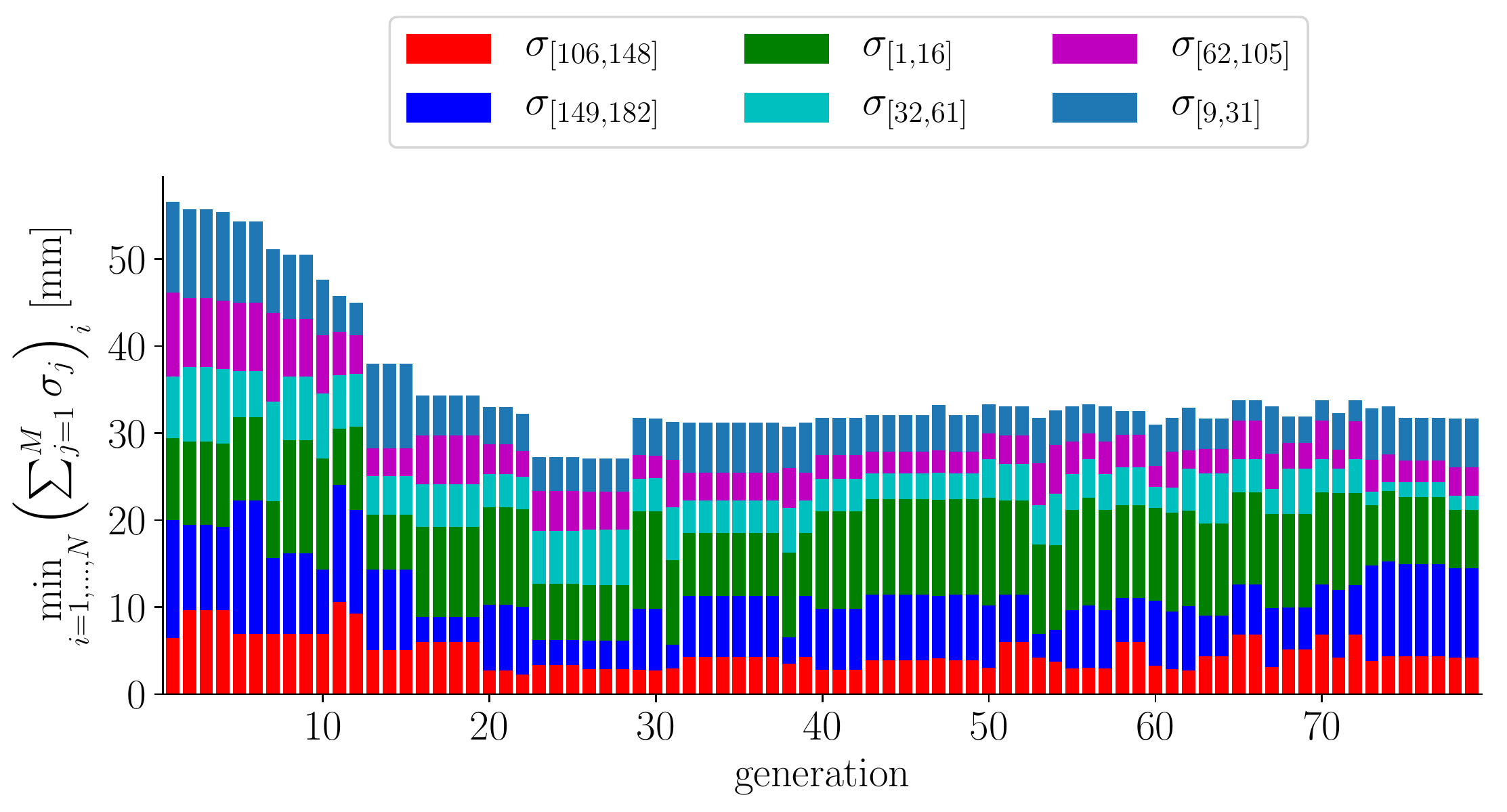}
    \caption{Evolution of the best individual during the multi-objective optimization.\ The best
    individual of a generation is identified by the smallest sum of objectives $\sigma_{j}$ with
    $j\in [1, ..., M]$ and $M$ the number of objectives.\
    The best individual was contained first in generation 26.\ The minimization is over
    all $N=8062$ individuals per generation.\ The label $\sigma_{[l,u]}$
    indicates an objective for the turns in the range $[l, u]$.}
    \label{fig:test_71_evolution_of_objectives}
\end{figure}

The optimization consisted of several independent runs with initially large bounds for each
design variable.\ These bounds were narrowed according to the best individual.\
A best individual per generation is defined as the smallest sum of all $M$ objectives $\sigma_j$, i.e.\
\begin{equation*}
\min\limits_{i=1,...,N}\left(\sum_{j=1}^{M}\sigma_j\right)_{i}.
\end{equation*}
We only show the evolution of the last optimization in \Figref{fig:test_71_evolution_of_objectives}
that was stopped after \num{79} generations since after \num{26} generations no significant error
reduction was observed.\

The objective values of the best individual over all generations are summarized in
\Tabref{tab:error_best_individual_optimizer}.\ According to \Figref{fig:turn_separation_rri2}
the smallest turn separation at RRI2 is $\SI{18}{mm}\gg\sigma_{[1,16]}=\SI{6.38}{mm}$,
where the symbol $\sigma_{[l,u]}$ indicates a single objective for the turns $l$ to $u$.\
Therefore, the deviation to the measurement is less than half a turn for the maximum
absolute error.\ In case of RRL the difference is also always below the turn separation
(cf.\ \Figref{fig:turn_separation_rrl1}).\

\begin{table}[htp]
    \caption{Result of best individual obtained by optimization using the
    $l_\infty$-error norm for each objective.\ The label $\sigma_{[l,u]}$
    indicates an objective for the turns in the range $[l, u]$.}
    \label{tab:error_best_individual_optimizer}
    \begin{ruledtabular}
    \begin{tabular}{lcl}
        Objective        & $l_{\infty}$-error & Probe \\
        $\sigma_{[l,u]}$     & (\si{mm})     & \\
        \midrule 
        $\sigma_{[1,16]}$    & \num{6.38} & RRI2\\
        $\sigma_{[9,31]}$    & \num{3.76} & RRL\\
        $\sigma_{[32,61]}$   & \num{6.34} & RRL\\
        $\sigma_{[62,105]}$  & \num{4.39} & RRL\\
        $\sigma_{[106,148]}$ & \num{2.91} & RRL\\
        $\sigma_{[149,182]}$ & \num{3.27} & RRL\\
    \end{tabular}
    \end{ruledtabular}
\end{table}

\subsection{Local search}
\label{sec:local_search}
While the genetic algorithm can in principle search a large variable space effectively,
it was not able to find a nearby better solution in a reasonable amount of time.\
We suspect this is due to high sensitivity of the design variables near the optimum, the heuristics of
genetic algorithms and the large dimensionality of the design variable space.\
Therefore, once the best individual from the genetic algorithm was selected,
a local search around this individual was done to find the optimum.\
The chosen local search involved changing a single parameter value iteratively.\
This approach reduced the turn pattern error significantly.\

\begin{figure}[htp]
     \centering
     \includegraphics[width=\columnwidth]{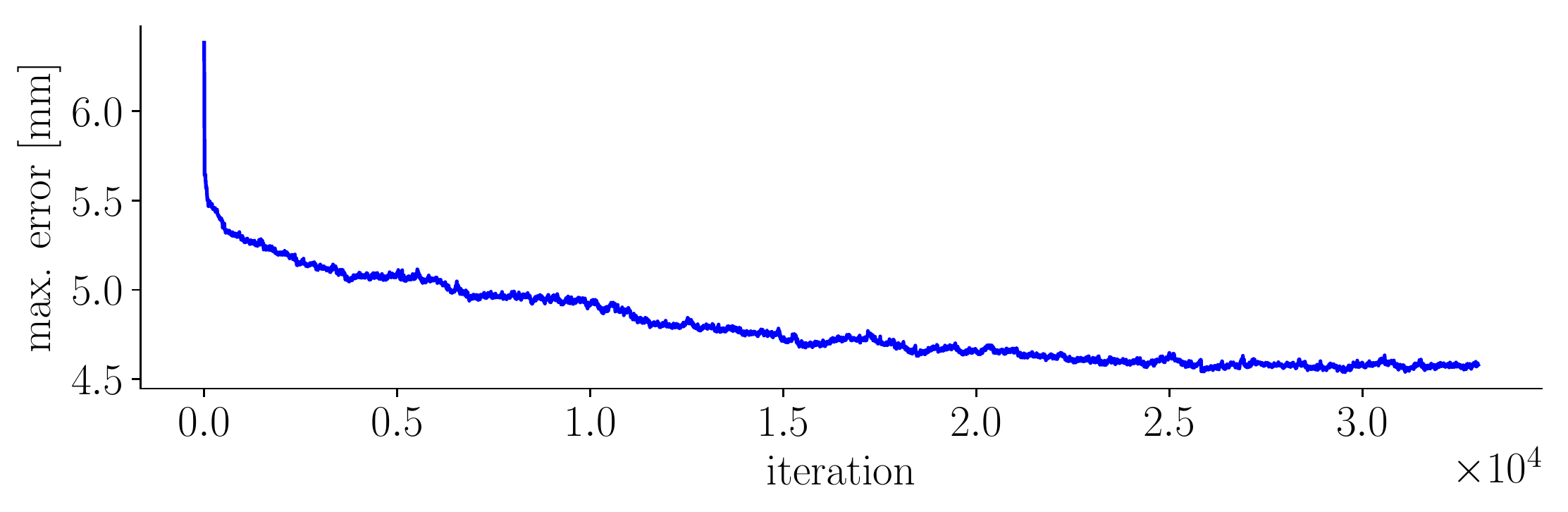}
     \caption{Evolution of the $l_{\infty}$-error between measurement and
     simulation during the local search with the best individual obtained
     by the MOGA as starting point.}
     \label{fig:scan_max_error}
\end{figure}

Defining a good metric for the search was crucial since the iterative search is likely to stop in a local optimum.\
To avoid local optima several norms were used simultaneously,
namely the maximal error ($l_{\infty}$-error),
the second largest error, the third largest error, 
and a weighted $l_2$-error where the RRI2 turns were weighted equally to the RRL turns.\
A parameter was allowed to change when there was an improvement in any of the norms while not worsening the other norms significantly
(\SI[round-precision=2]{0.01}{mm} for the $l_{\infty}$-error).\
This is equivalent to a multi-objective optimization.\
In \Figref{fig:scan_max_error} the $l_{\infty}$-error is shown during the iterative search.\
It can be seen that there was a significant improvement in the beginning, reducing the error from more than \SI{6}{mm} to less than \SI{5}{mm}.\
It can also be seen that the error occasionally increases which avoids the local optima.\
Once no improvement in the $l_{\infty}$-error over \num{30000} iterations was observed the local search was stopped.\

\begin{figure}[htp]
     \centering
     \includegraphics[width=\columnwidth]{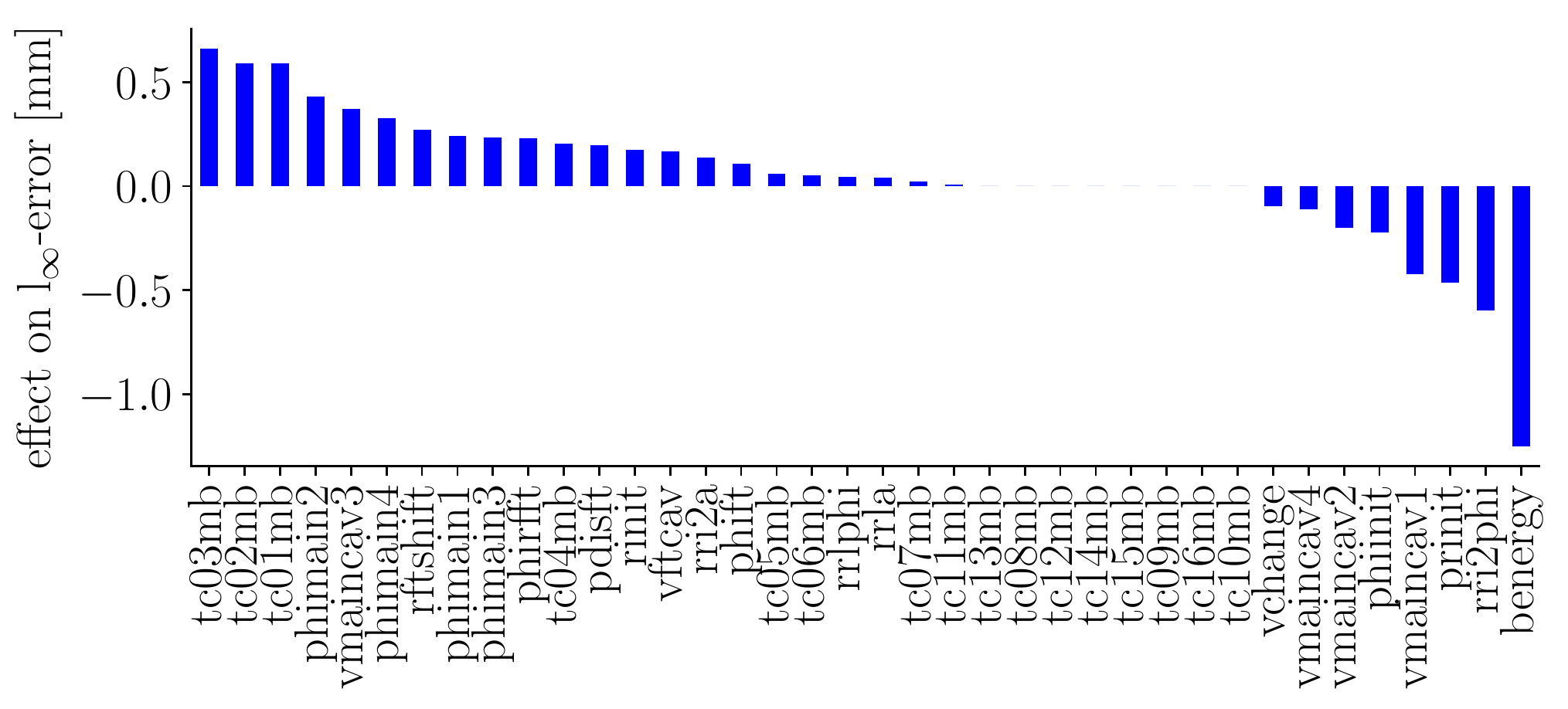}
     \caption{Total effect during the local search on the $l_{\infty}$-error per design variable.}
     \label{fig:scan_max_sum}
\end{figure}

\begin{figure}[htp]
     \centering
     \includegraphics[width=\columnwidth]{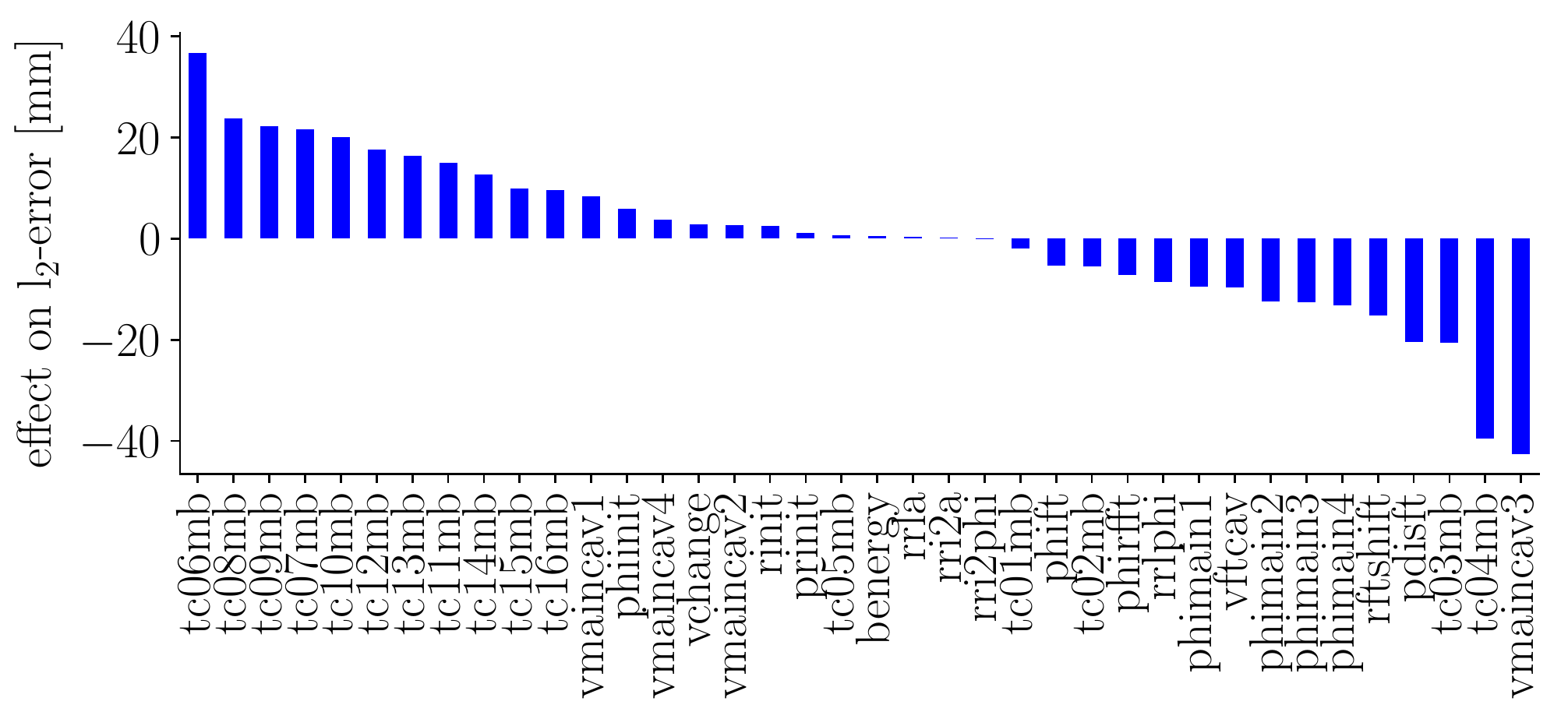}
     \caption{Total effect during the local search on the $l_{2}$-error per design variable.}
     \label{fig:scan_l2_sum}
\end{figure}

In \Figref{fig:scan_max_sum} and \Figref{fig:scan_l2_sum} the effect on the $l_{\infty}$-error and $l_{2}$-error per design variable is shown.\
It can be seen that as expected the trim coils generally improve the $l_{2}$-error, 
while the first trim coils and RF improve the $l_{\infty}$-error.\
The explanation for the latter is that the largest mismatch during the scan was often in one of the first turns.\

%
%

    
    \section{Discussion}
\label{sec:discussion}
\begin{figure}[htp]
    \centering
    \includegraphics[width=\columnwidth]{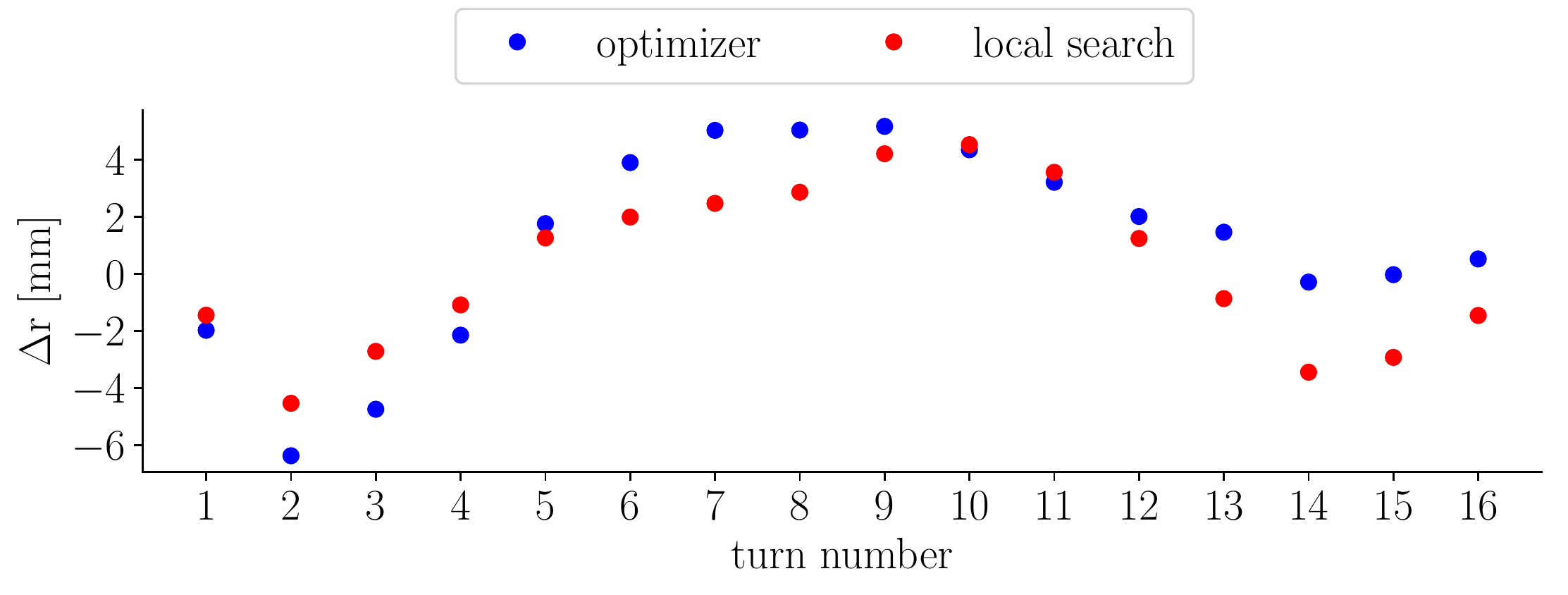}
    \caption{Error of the turn radius at RRI2 between measurement and simulation
    of the best individual obtained by multi-objective optimization and local search.}
    \label{fig:peak_difference_optimizer_rri2}
\end{figure}

\begin{figure}[htp]
    \centering
    \includegraphics[width=\columnwidth]{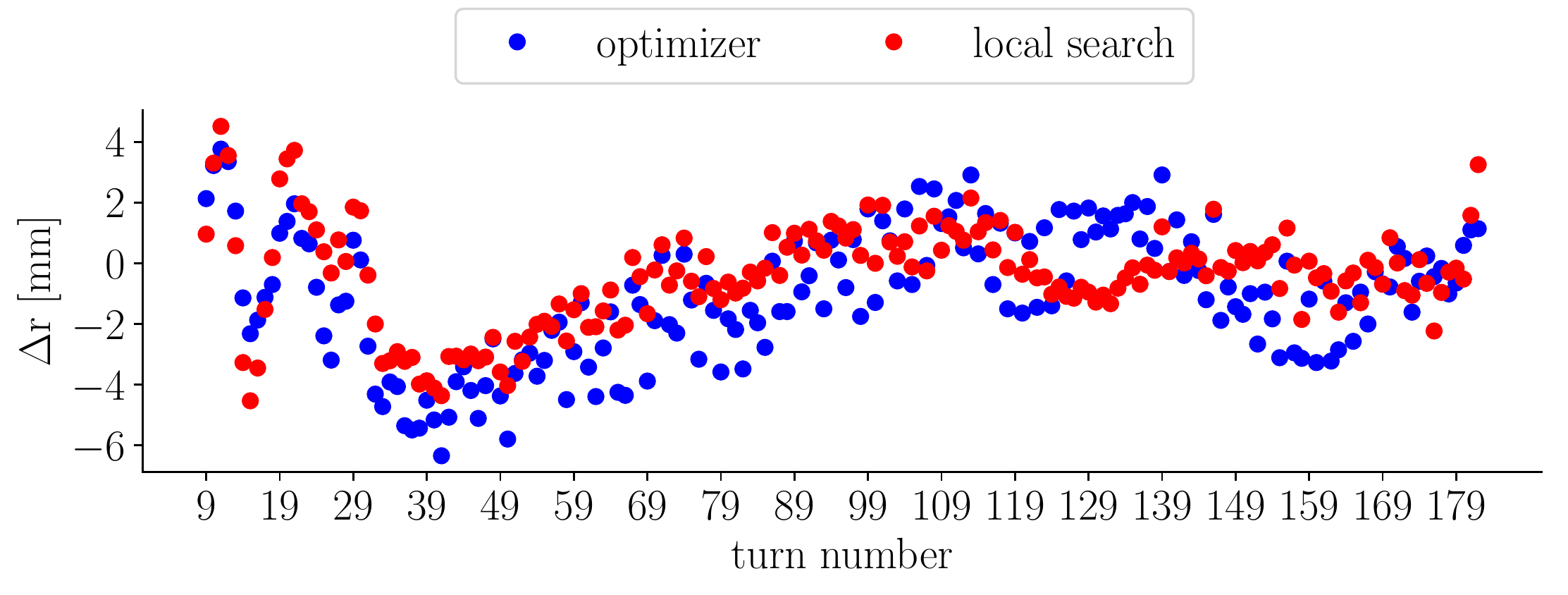}
    \caption{Error of the turn radius at RRL between measurement and simulation
    of the best individual obtained by multi-objective optimization and local search.}
    \label{fig:peak_difference_optimizer_rrl1}
\end{figure}

The starting point of the local search was the best individual of
the MOGA as explained in \Secref{sec:local_search}.\ This
additional step could reduce the error spread and maximum
absolute error (at turn \num{2}) compared to the measurement
(cf.\ \Tabref{tab:error_optimizer_and_scan}).\ The error of the turn
radius of both methods is shown in
\Figref{fig:peak_difference_optimizer_rri2} and
\Figref{fig:peak_difference_optimizer_rrl1}.\

The result of the single particle local search
is verified with two multi particle tracking simulations at \SI{88}{\micro\ampere}
having \num{360000} macro particles and either space charge (i.e.\ FFT Poisson solver)
switched on or off.\
The multi particle tracking (no space charge) changes the error compared to the
measurement only slightly.\ The maximum absolute error is increased by \SI{0.1}{mm} in
comparison to the single particle simulation.\ The mean absolute error (MAE) and mean squared error
rise only by $+\SI{0.061789474}{mm}$ and $+\SI{0.1813}{mm\squared}$ respectively.\

A multi particle tracking simulation with space charge doesn't change the turn pattern perceptibly
(cf.\ \Tabref{tab:error_no_sc_vs_meas_or_sc}).\ The $l_{\infty}$-error between
both multi particle simulations differs by \SI[round-precision=2]{0.0500000000002}{mm} and MAE
\SI[round-precision=2]{0.00415789473685}{mm}.\ These observations confirm the model assumptions to
neglect space charge and to use a single particle only in order to match the
turn pattern.\

An estimation of the error due to the measurement and model simplifications is
given in \Tabref{tab:error_source}.\ The systematic error is \SI{3.9}{mm} which
is comparable to \SI{4.54}{mm} of the local search (cf.\ 
\Tabref{tab:error_optimizer_and_scan}).\ The MAE also differs only by 
\SI{0.3}{mm}.\ The difference of the MSE is, however, \SI{2.5}{mm}.\

\begin{table}[htp]
    \caption{Maximum absolute error ($l_{\infty}$-norm), mean absolute error (MAE)
    and the mean squared error (MSE) of the best individual of the optimizer
    and local search compared to the measurement.\
    In both cases the maximum error is at turn \num{2}.}
    \label{tab:error_optimizer_and_scan}
    \begin{ruledtabular}
    \begin{tabular}{lccc}
        Method & $l_{\infty}$-norm & MAE & MSE \\
               & (\si{mm}) & (\si{mm}) & (\si{mm\squared}) \\
        \midrule
        optimizer       & \num{6.38}
                        & \num{2.02905263158}
                        & \num{6.30194105263} \\
        local search    & \num{4.54}
                        & \num{1.40057894737}
                        & \num{3.40449526316} \\
    \end{tabular}
    \end{ruledtabular}
\end{table}

\begin{table}[htp]
    \caption{Maximum absolute error ($l_{\infty}$-norm), mean absolute error (MAE)
    and the mean squared error (MSE) of the measurement or multi particle
    tracking simulation including space charge to the multi particle tracking simulation
    neglecting space charge.}
    \label{tab:error_no_sc_vs_meas_or_sc}
    \begin{ruledtabular}
    \begin{tabular}{lccc}
        Comparison to & $l_{\infty}$-norm & MAE & MSE \\
                      & (\si{mm}) & (\si{mm}) & (\si{mm\squared}) \\
        \midrule
        measurement       & \num{4.64}
                          & \num{1.46236842105}
                          & \num{3.58579526316} \\
        space charge      & \num{0.0500000000002}
                          & \num{0.00415789473685}
                          & \num{0.0000563157894741} 
    \end{tabular}
    \end{ruledtabular}
\end{table}

\begin{table}[htp]
    \caption{Estimation of the lower bound of the error due to model simplifications and
    measurement inaccuracies.}
    \label{tab:error_source}
    \begin{ruledtabular}
    \begin{tabular}{lccc}
        Error source & $l_{\infty}$-norm & MAE & MSE \\
                     & (\si{mm}) & (\si{mm}) & (\si{mm\squared}) \\
        \midrule
        measurement (cf.\ \Figref{fig:peak_diffs53-55}) & \num{2.2}
                                                        & \num{0.397988505747}
                                                        & \num{0.302040229885} \\
        multi particle (no space charge)                & \num{0.1} & \num{0.1} & \num{0.2} \\
        space charge effect                             & \num{0.0500000000002} & \num{0.00415789473685}
                                                        & \num{0.0000563157894741} \\ 
        step size (cf.\ \Secref{sec:turn_pattern_matching}) & \num{1.48} & \num{0.560344827586} & \num{0.449387356322} \\
        \midrule
        sum                                             & \num{3.9} & \num{1.1} & \num{0.9} \\
    \end{tabular}
    \end{ruledtabular}
\end{table}

    \section{Conclusion}
A realistic simulation of existing cyclotrons heavily depends on measured data
and the accurate parameter specification of the machine.\ Furthermore, the precision
of the numerical models of devices such as RF cavities, radial probes
as well as the time discretization have an impact on the result.\ While the
latter is improved by a higher resolution at the expense of longer time to solutions
of the simulation or a more accurate time integrator, numerical models of devices are
only enhanced by better methods.\

A new, more realistic
trim coil model in the beam dynamics code \opal{} that supersedes the model developed
in \cite{PhysRevSTAB.14.054402} was presented.\ The model uses rational functions to describe the
shape of the trim coil field in radial direction.\ Although the model was applied to
the PSI Ring cyclotron, it is applicable to any circular type of machine.\

Thanks to the flexibility of the model, it could be used to match the turn pattern of the simulation
with measurements of a centered beam in the PSI Ring cyclotron.\
In order to match all 182 turns a multi-objective optimization
was applied with a parameter space including 16 trim coils and
32 other design variables, such as beam injection energy, RF cavity voltages and various element positions.\ The full
list of design variables is given in \Tabref{tab:dvars}.\
This process was complemented with a
local search starting from the best individual of the MOGA.\ That way
the absolute error between simulation and measurement could be reduced
to at most \SI{4.54}{mm}.\ Despite several simplifications
on the optimization procedure, the multi particle tracking without space charge
verified the matching of the single particle tracking.\ 
Nevertheless, the numerical model can be improved further, 
especially the azimuthal location of the trim coil field could be enhanced and, 
if possible, a 3-dimensional representation is aimed.\
In addition future work will include the matching of a non-centered beam and the beam profile.\

The proposed approach of multi-objective optimization is unique to this kind
of problem and might be used as a guideline for future projects.\
One of them is the DAE$\delta$ALUS project \cite{doi:10.1063/1.4802375, doi:10.1063/1.4826879},
a proposed search for
CP-violation in the neutrino sector.\ The DAE$\delta$ALUS Superconducting Ring Cyclotron (DSRC)
shares many similarities with the PSI Ring cyclotron.\ Future design
studies will benefit greatly from the newly developed methods that were presented here.

    \begin{acknowledgments}
The authors thank W.~Joho, M.~Humbel, R.~D\"olling and H.~Zhang for their
helpful discussions about trim coils and probe measurements.\ The contribution
of D.~Winklehner regarding the DAE$\delta$ALUS is gratefully acknowledged as well.\
Furthermore, we appreciate the expertise of M.~Kranj\v{c}evic regarding the
multi-objective optimization using evolutionary algorithms.\ All optimizations
were performed using the HPC resources provided by the Swiss National
Supercomputing Centre (CSCS).\ This project is funded by the Swiss National
Science Foundation (SNSF) under contract number 200021\_159936.
\end{acknowledgments}

    \appendix*
    \section{Design Variables}
\label{app:results}

\begin{table}[htp]
    \caption{Design variable abbreviations and their meaning.}
    \label{tab:dvars}
    \begin{ruledtabular}
    \begin{tabular}{>{\raggedright}p{3cm}cp{4.5cm}}
        Design variable & Unit & Meaning \\
        \midrule
        benergy                     & \si{GeV}              & injection beam energy \\
        pdisft                      & \si{mm}               & displacement of flat top's axis from global center \\
        pdismain1 - pdismain4       & \si{mm}               & displacement of main cavity's axis from global center \\
        phift                       & \si{\deg}             & flat top cavity angle w.r.t.\ global coordinate system \\
        phiinit                     & \si{\deg}             & injection angle of beam \\
        phimain1 - phimain4         & \si{\deg}             & main cavity's angle w.r.t.\ the center line of sector magnet 1 \\
        phirfft                     & \si{\deg}             & phase of flat top \\
        prinit                      & \si{\beta\gamma}      & injection radial momentum \\
        rftshift                    & \si{mm}               & flat top cavity displacement in radial direction\\
        rinit                       & \si{mm}               & injection radius w.r.t.\ the global coordinate system \\
        rmainshift1 - rmainshift4   & \si{mm}               & main RF cavity displacement in radial direction\\
        rri2a                       & \si{mm}               & $a$ of \Eqref{eq:probe_positioning} for RRI2 \\
        rri2phi                     & \si{\deg}             & $\varphi$ of \Eqref{eq:probe_positioning} for RRI2 \\
        rri2shift                   & \si{mm}               & start position of RRI2 in radial direction ($> 0:$ outwards) \\
        rrla                        & \si{mm}               & $a$ of \Eqref{eq:probe_positioning} for RRL \\
        rrlphi                      & \si{\deg}             & $\varphi$ of \Eqref{eq:probe_positioning} for RRL \\
        rrlshift                    & \si{mm}               & start position of RRL in radial direction ($> 0:$ outwards) \\
        tc01mb - tc16mb             & \si{T}                & trim coil maximum magnetic field \\
        vchange                     & \si{MV}               & extra RF voltage change of main cavities (in total) \\
        vftcav                      & \si{MV}               & RF voltage on flat top cavity \\
        vmaincav1 - vmaincav4       & \si{MV}               & RF voltage on main cavity 1 - 4 \\
    \end{tabular}
    \end{ruledtabular}
\end{table}

    \clearpage
    
    \bibliographystyle{apsrev4-1}
%

\end{document}